\documentclass[aps,prb,twocolumn,groupedaddress,showpacs]{revtex4-1}
\usepackage{epsfig}
\usepackage[dvipsnames,usenames]{color}
\usepackage[normalem]{ulem}
\usepackage{amsmath}

\emergencystretch=\maxdimen
\begin{document}
\title{Magnetic and metal-insulator transitions in coupled
spin-fermion systems}
\author{R.~Mondaini$^{1,2}$, T.~Paiva$^{3}$,  and R.T.~Scalettar$^{2}$}

\affiliation{$^1$Physics Department, The Pennsylvania State University, 104
Davey Laboratory, University Park, Pennsylvania 16802, USA}

\affiliation{$^2$Physics Department, University of California, Davis,
California 95616, USA}

\affiliation{$^3$Instituto de F\'isica, Universidade Federal do Rio de
Janeiro Cx.P. 68.528, 21941-972 Rio de Janeiro RJ, Brazil}

\begin{abstract}
We use quantum Monte Carlo to determine the magnetic and transport
properties of coupled square lattice spin and fermionic planes as a
model for a metal-insulator interface.  Specifically, layers of Ising
spins with an intra-layer exchange constant $J$ interact with the
electronic spins of several adjoining metallic sheets via a coupling
$J_H$.  When the chemical potential cuts across the band center, that
is, at half-filling, the N\'eel temperature of antiferromagnetic ($J>0$)
Ising spins is enhanced by the coupling to the metal, while in the
ferromagnetic case ($J<0$) the metallic degrees of freedom reduce the
ordering temperature.  In the former case, a gap opens in the fermionic
spectrum, driving insulating behavior, and the electron spins also
order.  This induced antiferromagnetism penetrates more weakly as the
distance from the interface increases, and also exhibits a non-monotonic
dependence on $J_H$.  
For doped lattices an interesting charge disproportionation occurs where
electrons move to the interface layer to maintain half-filling there.
\end{abstract}

\maketitle

\section{Introduction}

Over the last several decades an extensive literature has developed
describing Monte Carlo simulations of both localized (e.g.~Heisenberg)
and itinerant (e.g.~Hubbard) models of quantum magnetism.  An important
subset of these studies has considered situations where the exchange
constants $J_\alpha$ or electron repulsion $U_\alpha$
can take on multiple values, with the
attendant possibility of quantum phase transitions as the ratio of these
energy scales is altered.  For example, in the case of the one-fifth
depleted square lattice model of CaV$_4$O$_9$, the quotient $J/J'$ of
the exchange constants on the two different vanadium  bonds tunes the
associated Heisenberg hamiltonian from a disordered dimer phase, to
N\'eel order, and then back to a disordered plaquette
phase\cite{troyer96}, lending an understanding of spin gapped behavior
in this material.  Likewise, bilayer Heisenberg\cite{sandvik94} and
Hubbard\cite{scalettar94} models have a singlet to N\'eel transition
depending on the ratio of values of the inter- and intra-plane energies.

In addition to describing systems in which long range order can be
destroyed, multiple $J_\alpha$ and $U_\alpha$ can also give rise to
transitions between different ordered states, such as charge density
versus spin density wave patterns.  Simulations of models with several
interaction energy scales are especially relevant to heterostructures,
where the growth of distinct sheets of the same, or different materials,
offers the possibility of tuned magnetic properties.

In this paper we present a quantum Monte Carlo investigation of a mixed
localized-itinerant magnetic model in which we couple a 2D square layer
of Ising spins to several metallic planes.  Our interest is both in how,
potentially, the additional fluctuations of the free electrons suppress
the Ising transition temperature and on how the magnetic layer initiates
order amongst the free fermions.  We also explore whether the coupling
of the metal to the localized spins can open a gap in the electronic
spectrum, driving a metal to insulator transition, and the penetration
depth of the magnetic order into the metal.  Our work is related to
simulations of multilayer Hubbard models in which the on-site
interaction $U$ can distinguish metallic from magnetic
layers\cite{euverte12}.  However, by treating the correlated layers as
classical, localized spins we are able to explore a greater range of
parameter space, and, in particular, to go to lower temperatures away
from half-filling of the metallic band where a sign problem would
otherwise prevent simulations.  

This idea of coupling classical spins to itinerant electrons has been
extensively used, e.g.~in multiband models of the manganites and iron
pnictides where the sign problem similarly precludes treating fully
quantum mechanical models \cite{dagotto01,lv10,yin10}.  Numerical
approaches to these models allow easier access to dynamical behavior and
hence greater possibility of contact with spectroscopy and neutron
scattering experiments\cite{liang12} than do direct path integral
treatments of many-electron systems which require a difficult analytic
continuation to get real time information.

A number of recent 
experiments have examined electronic reconstruction at the 
interface of different transition metal oxides using
scanning tunneling microscopy with high spatial and energy 
resolution.  Some of these experiments
focus on interfaces of paramagnetic metals and 
antiferromagnetic insulators \cite{takahashi01,freeland10,tan13}.
The Hamiltonian we consider here,
in which tight binding layers couple to classical, localized spins,
is the most simple model of such a situation,
and will clearly require
considerable refinement before being able to make any sort of
quantitative contact.  Nevertheless, it can lend a first qualitative
insight into the sort of trends one might expect, e.g.~for magnetic
order.  Moreover, 
the study of fluctuating classical spins coupled to fermionic degrees of freedom
has recently been suggested as a generally promising
approach to move beyond mean-field treatments of interacting
electron systems \cite{dagotto14},
providing further motivation for this work.

The remainder of this paper is organized as follows: In
Sec.~\ref{sec:model} we write down the fermion-Ising hamiltonian along
with a brief summary of the numerical methods employed, and definitions
of the observables which characterize the phases.  Section
\ref{sec:results} describes the results when the Fermi level
is at the band center, first for the case of
antiferromagnetic (AF) Ising spins and then for ferromagnetic
coupling, followed by a discussion
(Sec.~\ref{sec:resultsdoped}) of the effect of doping away from half-filling.  
A particular interesting charge disproportionation is shown
to occur where metallic layers become unequally populated to allow
for an optimal magnetic response.
Section \ref{sec:conclusion} presents a conclusion
and some future directions of research.

\section{Model}
\label{sec:model}

\begin{figure}[t]
\vspace{+0.5cm}
\epsfig{figure=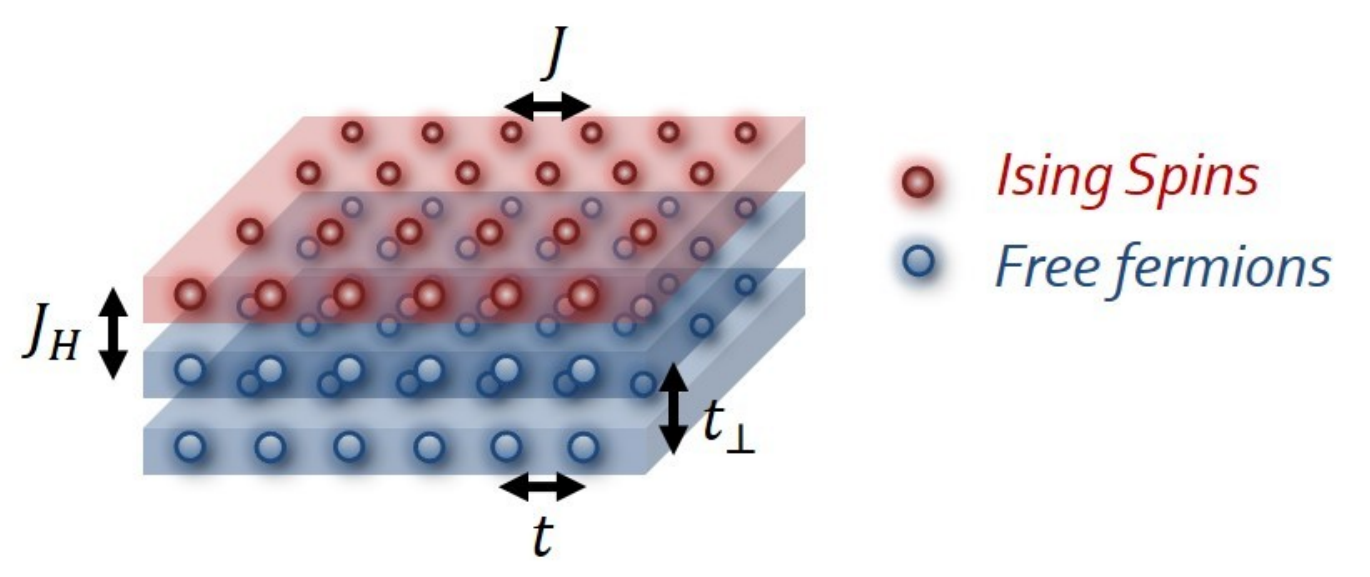,width=9.0cm,angle=-0,clip}
\caption{(Color online)
Lattice geometry for the fermion-Ising model.  A single layer of
Ising spins residing on a 2D square lattice is superposed
on several layers of noninteracting fermions.
The nearest-neighbor 
spin-spin interaction between the free fermions of layer
$\ell=0$ and the Ising spins is 
proportional to the parameter $J_H$.
\label{fig:cartoon}
}
\end{figure}

We consider the hamiltonian
\begin{align}
\hat H = -&t \sum_{\langle ij\rangle \ell,\sigma} \big( \, 
c_{i \ell \sigma}^{\dagger} c_{j \ell \sigma}^{\phantom{\dagger}}
+ c_{j \ell \sigma}^{\dagger} c_{i \ell \sigma}^{\phantom{\dagger}}
\, \big)-\mu \sum_{i \ell,\sigma}  n_{i \ell\sigma}
\nonumber \\
 -&t_\perp \sum_{i \langle \ell \ell'\rangle,\sigma} \big( \, 
c_{i \ell \sigma}^{\dagger} c_{i \ell' \sigma}^{\phantom{\dagger}}
+ c_{i \ell' \sigma}^{\dagger} c_{i \ell \sigma}^{\phantom{\dagger}}
\, \big)
\nonumber \\
-& J_H\sum_{i} {s}^z_{i,\ell=0}
 \, {S}_i 
+J\sum_{\langle ij\rangle}{S}_i \, {S}_j ,
\label{eq:hamiltonian}
\end{align}
where $c_{i\ell\sigma}^{\dagger}\,(c_{i\ell\sigma}^{\phantom{\dagger}})$
are creation(destruction) operators for fermions of spin $\sigma$ on
lattice site $i$ of layer $\ell=0,1,\cdots,N_{\rm layer}-1$.  
Our convention is that layer $\ell=0$ is adjacent to the classical
spins.
The intra-layer hopping $t$ is on
nearest-neighbor sites (denoted by $\langle ij \rangle$) of each layer
$\ell$; the inter-layer hopping between neighboring fermionic layers
$\langle \ell \ell' \rangle$ is $t_\perp$, and the density of fermions
is tuned by the chemical potential $\mu$.  The geometry of each layer is
that of a 2D square lattice of linear length $L$.  The remaining
degrees of freedom are Ising spins which populate a single
layer\cite{footnote1} and are coupled by exchange constant $J$.  The
Ising spins interact with the $z$ component of the fermion spin
 ${s}^z_{i,0}= n_{i,0, \uparrow} - n_{i,0, \downarrow}$
in the interface layer $\ell=0$, via a second exchange constant, $J_H$.
The lattice geometry is sketched in Fig.~\ref{fig:cartoon}.  We choose
periodic boundary conditions in the planes, and open boundary conditions
in the direction perpendicular to the planes. 
Our results will be for two
metallic layers (i.e.~$N_{\rm layer}=2$), since, as we shall show, such
a situation already allows us to address many of the key questions
concerning the interface between a magnetic and a metallic layer.

We have chosen $|J|/t=0.2$ (both signs of $J$ will be studied) so that the 
temperature scale
for the development of correlations in the classical
spins is comparable to that in the metallic layer and, consequently, possible
competing phases are most readily discerned. There are different ways to 
understand this.
The most simple is to note that, if $J=t$, the 2D square
lattice Ising $T_c \sim 2.27 J$  is much higher than
typical temperature scales at which short range correlations get more robust 
for noninteracting fermions in a square lattice.
This is because
for the half-filled $U=0$ Hubbard Hamiltonian,
short range antiferromagnetic correlations corresponding to
the Fermi wavevector is $\mathbf{k}_F = (\pi,\pi)$,
do not onset until the temperature gets below $T \sim 0.25 t$.
Even when electron-electron interactions, which
are not considered here, are turned on, 
nearest neighbor spin correlations do not begin to grow 
substantially
until $T \sim 0.5 t$
(for the $U/t=4$
Hubbard model).  Thus in either case,
a choice $|J|/t \sim 0.2$ (Ising $T_c \sim 0.45 t)$ is required
to select
classical spin and fermionic spin ordering scales to be roughly equal.

An alternate to Eq.~\ref{eq:hamiltonian} would be to consider continuous
planar $\vec {\bf S} = (S_i^x,S_i^y)$ or Heisenberg $\vec {\bf S} =
(S_i^x,S_i^y,S_i^z)$ spins, with an $\vec {\bf S}_i \cdot \vec {\bf
S}_j$ spin-spin coupling between pairs of local spins, and $\vec {\bf
S}_i \cdot \vec {\bf s}_j$ spin-spin of local spin to fermion spin.\cite{dagotto01}  The restriction used here, to a single ($z$)
component, has been considered in other problems involving treating
electronic correlation, from mean field approaches\cite{chiesa13} to the
study of the $t$-$J_z$ model.\cite{zhang88}  The choice of Ising spins
also ensures a robust magnetic phase transition in which true long range
order occurs at finite $T_c$ in the spin plane.  This will be discussed
further in the conclusions.

It is worth noting several symmetries of the hamiltonian
Eq.~\ref{eq:hamiltonian}.  Consider first a combined particle-hole
transformation $\,c_{i\sigma}^{\phantom{\dagger}} \rightarrow (-1)^i
c_{i\sigma}^{\dagger}$ and inversion of the localized spins $\big(\,S_i
\rightarrow-S_i\,\big)$.  Here $(-1)^i$ denotes a staggered $\pm 1$
phase taking opposite values on the two sublattices of the bipartite
square lattice.  This transformation leaves each of the terms in the
hamiltonian-  the fermion kinetic energy, the Ising interaction, and the
local spin-fermion coupling invariant.  Thus, if $\mu=0$, the whole
hamiltonian is unchanged, and the lattice is 
half-filled ($\rho=1.0$).  

The finite temperature properties of the system can be obtained from its
partition function and associated expectation values.  The partition
function is,

\begin{equation}
Z=\sum_{S_i=\pm 1} 
e^{\beta J \sum_{\langle ij \rangle} S_i S_j} \cdot Z_{\rm f} (\{S\}),
\label{eq:Ztotal}
\end{equation}
where $Z_{\rm f}(\{S\})={\mathrm{Tr}} \, 
e^{-\beta({\hat H}_{{\rm f}\uparrow}
+{\hat H}_{{\rm f}\downarrow})}$ 
represents the grand-canonical partition function of the fermionic part
of the hamiltonian 
for a
particular Ising field configuration $\{S\}$. Since the hamiltonian
Eq.~\ref{eq:hamiltonian} is bilinear in the fermionic operators, each
$\hat H_{\rm f \sigma}$ can be written as the product of a vector of
creation operators, a real-valued matrix ${\cal M}^\sigma$,\cite{footnote2} and 
a vector
of destruction operators.  The fermion contributions to $Z$ can then
be expressed in terms of the eigenvalues $\lambda_j^\sigma$ of ${\cal
M}^\sigma$,
\begin{equation} 
Z_{\rm f}=\prod_{\sigma=\uparrow,\downarrow}\prod_{j}\left(1+e^{
\beta\lambda^\sigma_j}\right)  .
\end{equation}

From this expression, it is clear that the summand in
Eq.~\ref{eq:Ztotal} is positive definite and there is no ``sign
problem'' (for any $\mu$).  Of course, this is simply a consequence of
the fact that the spin field to which the fermions are coupled does not
vary in imaginary time, as it would, for example, if $\{S\} = S_{i\tau}$
were a Hubbard-Stratonovich field used to decouple a fermion-fermion
interaction.  The largest computational effort arises from diagonalizing
the two $N$x$N$ matrices ${\cal M}^\sigma$ for each update to the
configuration of the Ising spins $S_i$.

An alternate method to the direct matrix diagonalization used in the literature 
employs a representation of the density of states $\rho(\lambda)$ in
terms of Chebyshev polynomials \cite{motome99,alvarez05,cen06}.
The moments of $\rho(\lambda)$ are computed recursively in
a way that involves only sparse matrix-vector multiplications.
This approach
improves the scaling with system size to linear in $N$, 
at the cost of a significant
prefactor. It  also involves a (well-controlled) approximation which is
the truncation of the expansion at some maximum order.
Here we use exact diagonalization, as opposed to the Chebyshev method.

The results of the simulations presented below were obtained by
averaging over 5-10 independent simulations, each of which was composed
of 35,000 Monte Carlo sweeps of the Ising variables. Typically, the
linear lattice size $L$ was varied between $4 \le L \le 12$, selecting a
geometry with one Ising plane and stacked on top of two fermionic ones,
so that $N=2\,L^2$.

Expectation values of the Ising variables are averaged in the usual way
over the configurations generated in the course of the simulation.  For
example, to address directly if there is long range ferromagnetic order
in the Ising plane in the case $J<0$, we calculate the fourth order
Binder cumulant~\cite{Binder81}, $B_4(T)=\left(1-\langle
M^4\rangle/3\langle M^2\rangle ^2\right)$.  Here $M=1/N\sum_i S_i$ is
the magnetization per site.  When $J>0$ (the antiferromagnetic case) we
replace $M$ by the staggered magnetization, $1/N\sum_i (-1)^i S_i$.
Crossings of $B_4(T)$ obtained for different lattice
sizes determine the critical temperature for magnetic ordering of the
classical spins.  When the interaction $J_H$ between the Ising and
fermionic spins is nonzero, we expect a shift away from the 2D square
lattice Ising $T_c=2.269 \, |J|$.

Fermionic measurements like the kinetic energy, double occupancy, and
spin-spin correlations can be written in terms of combinations of the
single particle Green's functions,
$G_{ij}^\sigma=\langle  \, c^{\phantom{\dagger}}_{i\sigma}
c_{j\sigma}^{\dagger} \, \rangle = ({\cal M}_{ij}^\sigma)^{-1}$, for
every configuration of the classical spins.  The elements of
$G^{\sigma}$ are easily obtained from the diagonalization of ${\cal
M}^\sigma$ which is already in hand from the update of the spin variables.
Further details of the numerical algorithm for coupled classical
spin-fermion systems are contained in
Refs.~[\onlinecite{motome99,cen06,dagotto01}].

It is known from related simulations of Hubbard hamiltonians that 
fermions with no direct interaction $U$ have large finite size
effects:  the discrete (and often highly degenerate) $U=0$ energy levels
$E(k_x,k_y)$ are readily visible in measurements, especially dynamic
quantities like the density of states.  
Although the
$U=0$ metal considered here is coupled to classical spins, and hence
does have interactions, we still observe significant finite size
effects, especially in the metallic portions of the phase diagram.
We overcome this difficulty through the introduction of a small magnetic
field $B=\Phi_0/L^2$ along the direction perpendicular to the planes.
Here $\Phi_0$ is the magnetic flux quanta. With this choice, 
the intralayer hopping terms are 
changed by a Peierls-like phase factor $\big(\,t_{\bf ij}\rightarrow 
t \, {\rm exp} \,\big(\, \frac{2\pi i}{\Phi_0}
\int_\mathbf{i}^\mathbf{j}\mathbf{A}\cdot d\mathbf{l}\,\big)$.  We use
the Landau gauge in order to set the values of the vector potential
$\mathbf{A}$.  This procedure can be considered as an
improvement/generalization of ``boundary condition averaging"
\cite{gammel92,gros96,chiesa08}.  For a more complete description, see
Ref.~[\onlinecite{Assaad}]. Nevertheless, it is important to emphasize the distinction of this field, which couples to
the `orbital' motion of the electrons (i.e. their hopping)
from a Zeeman field coupling to spin which affects magnetic order.
The orbital field introduced here reduces finite size effects
by introducing an additional averaging over discrete allowed
momenta on a finite lattice.  The coupling to the classical spins,
on the other hand, produces a Zeeman field for the electrons,
whose role in ordering we will determine.

This reduction in the finite size effects is especially evident in the
single particle density of states, 
\begin{equation}
N(\omega)=\frac{1}{N}\, \mathrm{Im}\, \sum_{r} \sum_{j}
\frac{|U_{j,r}|^2} {\lambda_j-\omega-i\delta}.
\label{eq:dos}
\end{equation}
Here $U_{j,r}$ are the components of the eigenvectors
corresponding to the eigenvalue $\lambda_j$ of the matrices ${\cal
M}^\sigma$ defining the (quadratic) hamiltonian, and which now contain
the phase factors described above.  The outer sum averages all the
equivalent sites in order to recover translationally invariance.
Instead of displaying well-separated discrete delta-function peaks, even
for free fermions $N(\omega)$ becomes nearly continuous on relatively
small lattices, and has a form much closer to that of the thermodynamic
limit.\cite{Assaad}  In our hamiltonian, turning on $J_H$ further
reduces residual finite size effects.

\section{Results-  Half-filling}
\label{sec:results}

Because of Fermi surface nesting with vector ${\bf k}=(\pi,\pi)$, the
dominant magnetic instability of the half-filled square lattice Hubbard
hamiltonian is antiferromagnetic.  Indeed, the noninteracting
susceptibility $\chi_0(\pi,\pi)$ diverges as temperature $T \rightarrow
0$ so that, within the Random Phase Approximation (RPA), the ground
state exhibits AF order for any finite $U$.  Similarly, in the strong
coupling (Heisenberg) limit, the exchange interaction $J$ favors near
neighbor spins which are anti-aligned.  Since the $U=0$ fermion sheets
exhibit this strong AF preference, we expect a rather different response
to the coupling of an AF versus a F Ising plane to the metal.  We begin
with the AF case.

\begin{figure}[t]
\epsfig{figure=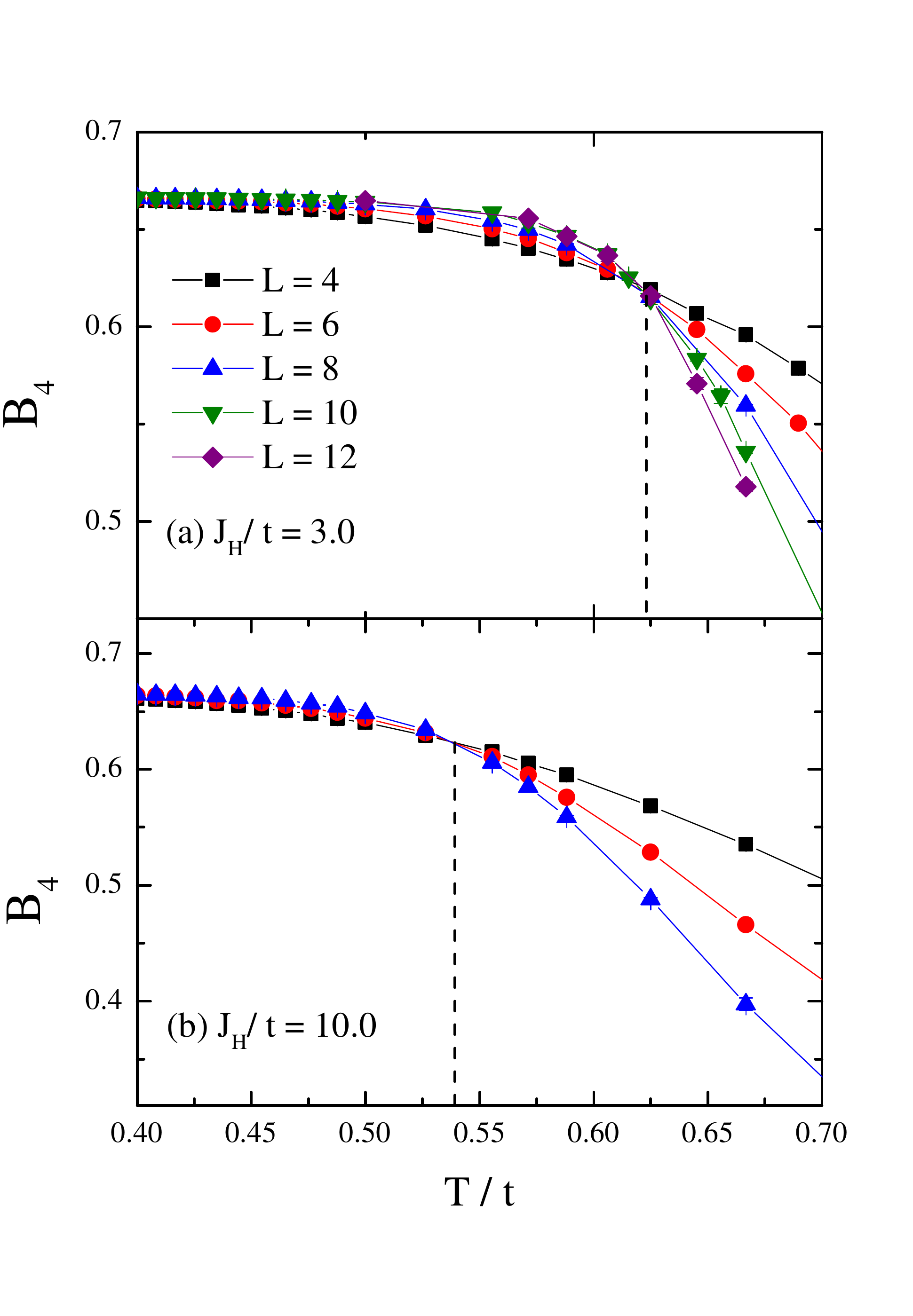,width=9.0cm,angle=-0,clip}
\caption{(Color online)
Crossing plot of the Binder ratio of an AF Ising sheet coupled with
$J_H=3t$ (top) and
$J_H=10t$ (bottom) to two metallic layers.  The interlayer hopping
between the fermion layers was set to $t_\perp = t$.
For $J_H=3t$,
the crossing occurs at $T/t \sim 0.62$, which is well above the
critical temperature of a free Ising sheet ($J_H=0$).
$T/t = (J/t) \cdot (T_c/J) = 0.2 \cdot 2.269 = 0.454$. The behavior of
$T_c$ with $J_H$ is nonmonotonic as the critical temperature for $J_H=10t$ drops to $T/t \sim 0.54$ (see Fig.~\ref{fig:T_cvsJ_H}).
\label{fig:Binder1}
}
\vspace{-0.5cm}
\end{figure}  

\begin{figure}[t]
\epsfig{figure=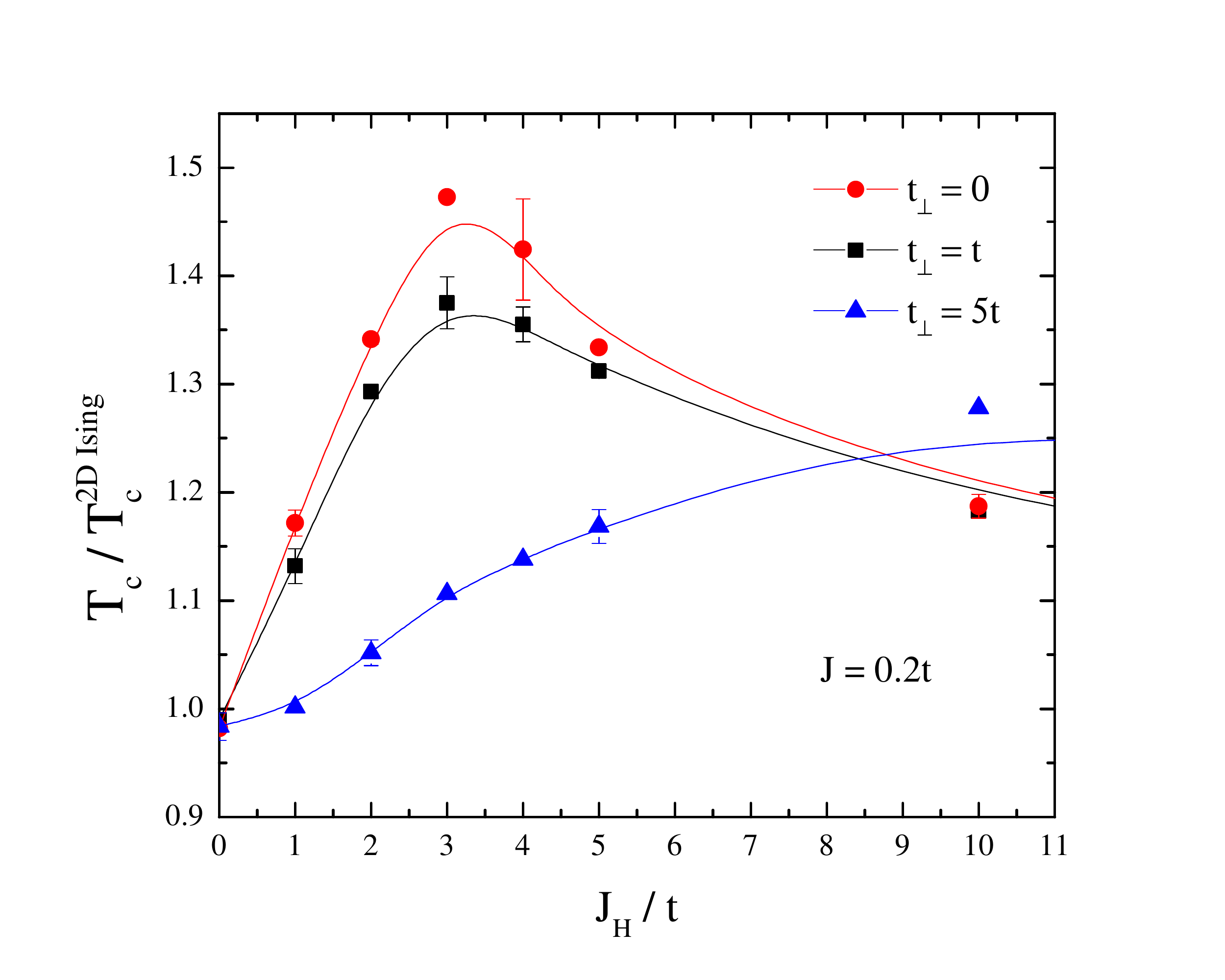,width=9.0cm,angle=-0,clip}
\caption{(Color online)
Critical temperature $T_c$ (normalized to the 2D Ising value)
for the magnetic transition in the Ising spin
plane as a function of the interaction $J_H$ with the metal.  Several
different choices of the hopping parameter $t_\perp$ connecting the two
metallic  planes are shown.  $T_c$ is obtained by the crossing of the
Binder ratio $B_4(T)$ for different lattice sizes.  
(See Fig.~\ref{fig:Binder1}.)
Coupling of the spin
layer to the metal enhances $T_c$.  The degree of enhancement is
strongest when the fermionic layers are most weakly coupled to each
other (small $t_\perp$).
Lines are guides only.
\label{fig:T_cvsJ_H}
}
\vspace{-0.5cm}
\end{figure}  

\subsection{The antiferromagnetic case}

Figure~\ref{fig:Binder1} shows the Binder ratio $B_4(T)$ for two
metallic planes of linear size $L$=4, 6, 8, 10 and 12 coupled in each case to a
single Ising plane of the same dimensions.  The interlayer hopping
between the fermionic layers is set to $t_\perp=t$ and the coupling
$J_H$ between the local spins and the fermions is $J_H=3 t$.  The Binder
ratios for the three lattice sizes cross at a common point, $T_c \sim
0.62$, representing a 36\% enhancement over the free spin plane result
$T_c \sim 2.27 J = 0.454$ for $J=0.2$.

Similar Binder crossing plots for other choices of $J_H$ and interlayer
hopping yield analogous transition temperatures, which are  shown in
Fig.~\ref{fig:T_cvsJ_H}.  
The enhancement in $T_c$ over that of an independent spin plane
is nontrivial, because there is a competition between the
additional entropy which results from fluctuations of fermionic
variables in the metallic plane and the tendency, noted above, towards
antiferromagnetism of the $U=0$ Hubbard model, due to Fermi surface
nesting.  Evidently, the latter tendency wins: $T_{\rm N\acute{e}el}$ is
enhanced. The universality class of the transition to an ordered
phase as the temperature is lowered remains an open question.
Our results are consistent with an Ising transition, and we believe
that is the most likely scenario, but the available system sizes do not allow us to draw any final conclusion.

\begin{figure}[t]
\epsfig{figure=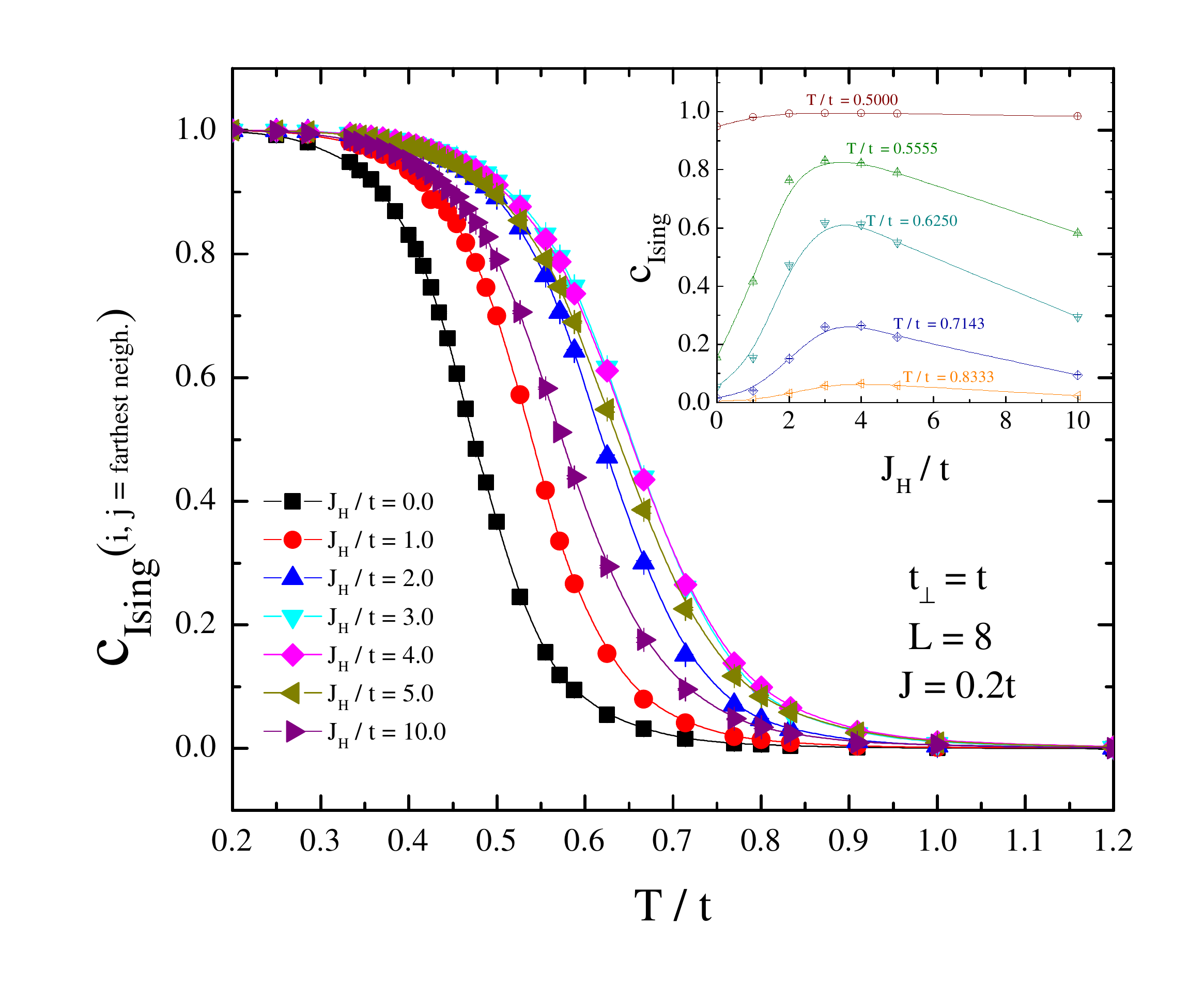,width=9.0cm,angle=-0,clip}
\caption{(Color online)
Temperature dependence of the Ising spin-spin correlation between most 
distant neighbors
in the lattice and several values of the interaction $J_H$ for $L=8$ with one 
Ising layer stacked over two fermion ones. 
These correlations display a similar trend to the critical temperature: they 
initially become more robust with
the coupling between the different planes; for 
large $J_H$, they return to values similar to the 
decoupled Ising model ($J_H=0$). The inset shows the same but now as a function of $J_H$ where the non-monotonic effects on the Ising spins correlations are unequivocal.
\label{fig:c_isingvsT}
}
\vspace{-0.5cm}
\end{figure}  

There are several additional interesting features in the data.
First, the enhancement in $T_c$ is non-monotonic in $J_H$.
The transition temperature reaches a maximum at $J_H/t \sim 3$ for 
both $t_\perp=0 $ and $t_\perp = t$.  Although data are not shown,
even for $t_\perp =5t $ the enhancement of $T_c$ comes back down 
at large $J_H$.
We note that the band structure of the two sheet Hubbard model is
$\epsilon(k_x,k_y) = -2 t \big(\,{\rm cos} \, k_x + {\rm cos} \, k_y
\, \big) \pm t_\perp$.  
The two bands overlap for $t_\perp < 4t$ and have a band gap $t_\perp -
4\,t$ otherwise.  Thus the choice $t_\perp = 5\,t$ represents the
coupling of an Ising spin layer to a band insulator rather than a metal.
Figure~\ref{fig:T_cvsJ_H} indicates that the magnetic response
of the Ising layer is qualitatively the same in the two situations
(metal with $t_{\perp}<4t$ or band insulator with $t_{\perp}>4t$),
although the response of a coupling to a band insulator produces less of
an effect, as might be expected.  This is likely due to the fact that
the bilayer Fermi surface is still nested with ${\bf k}=(\pi,\pi)$ for
$t_\perp > 4\,t$, even though the density of states at $E_F$ vanishes.

\begin{figure}[t]
\epsfig{figure=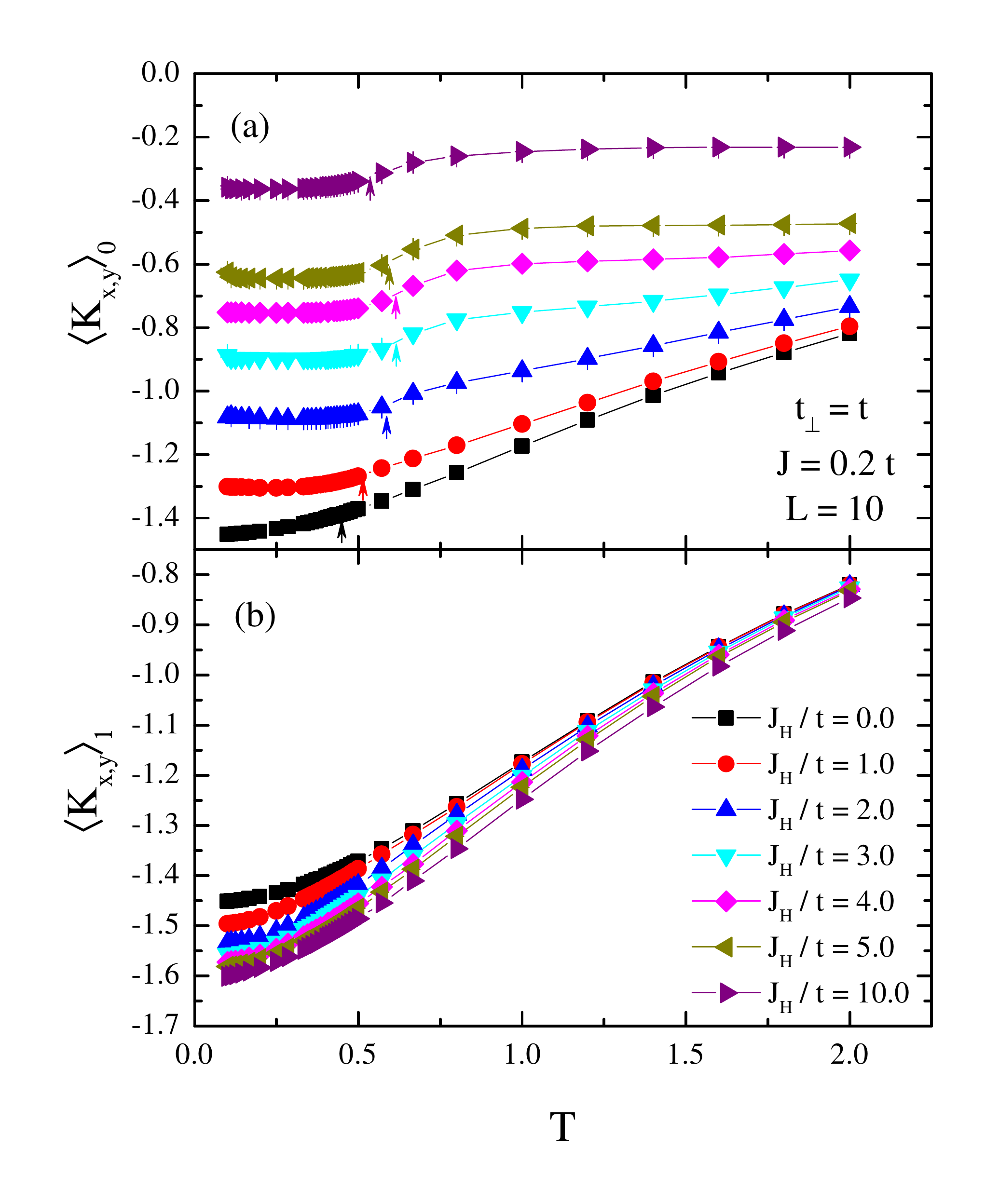,width=9.0cm,angle=-0,clip}
\caption{(Color online)
Intra-plane kinetic energy $\langle K\rangle$ as a 
function of temperature for different values of the interaction $J_H$.
Here $t_\perp=t$: In (a) the
fermionic plane $\ell=0$ directly in contact with the Ising spin layer; and
in (b) the more distant fermionic plane $\ell=1$. 
The trend with increasing $J_H$ is opposite in (a) and (b).
For $\ell=0$, 
the connection to the Ising spins reduces the kinetic energy 
at all temperatures.  For
$\ell=1$, the kinetic energy increases. The vertical arrows at panel (a) indicate the critical temperature below which the magnetic ordering takes place for the Ising spins (Fig.\ref{fig:T_cvsJ_H}).
\label{fig:kxyvsT}
}
\vspace{-0.5cm}
\end{figure}  

The non-monotonic behavior of $T_c$ with $J_H$ is reflected also in the
evolution of the farthest-neighbor intraplane spin correlation function.
$c(i,j)=\langle S_i S_j\rangle$.  Fig.~\ref{fig:c_isingvsT} shows
$c(i,j)$ vs.~$T/t$ for several values of $J_H$ at $t_{\perp}=t$ on a
8x8 lattice.  This quantity, which in the
thermodynamic limit would equal the square of the order parameter,
evolves rather sharply from zero to one as $T/t$ is lowered.  The
position where the switch in values occurs moves to larger $T/t$ as
$J_H$ changes from $J_H=0$ to $J_H/t \sim 3-4$, but then comes back down,
in agreement with the maximal $T_c$ in Fig.~\ref{fig:T_cvsJ_H}. The inset of Fig.~\ref{fig:c_isingvsT} displays the same quantity as a function of temperature and shows, unequivocally, this non-monotonic effect.

\begin{figure}[t]
\epsfig{figure=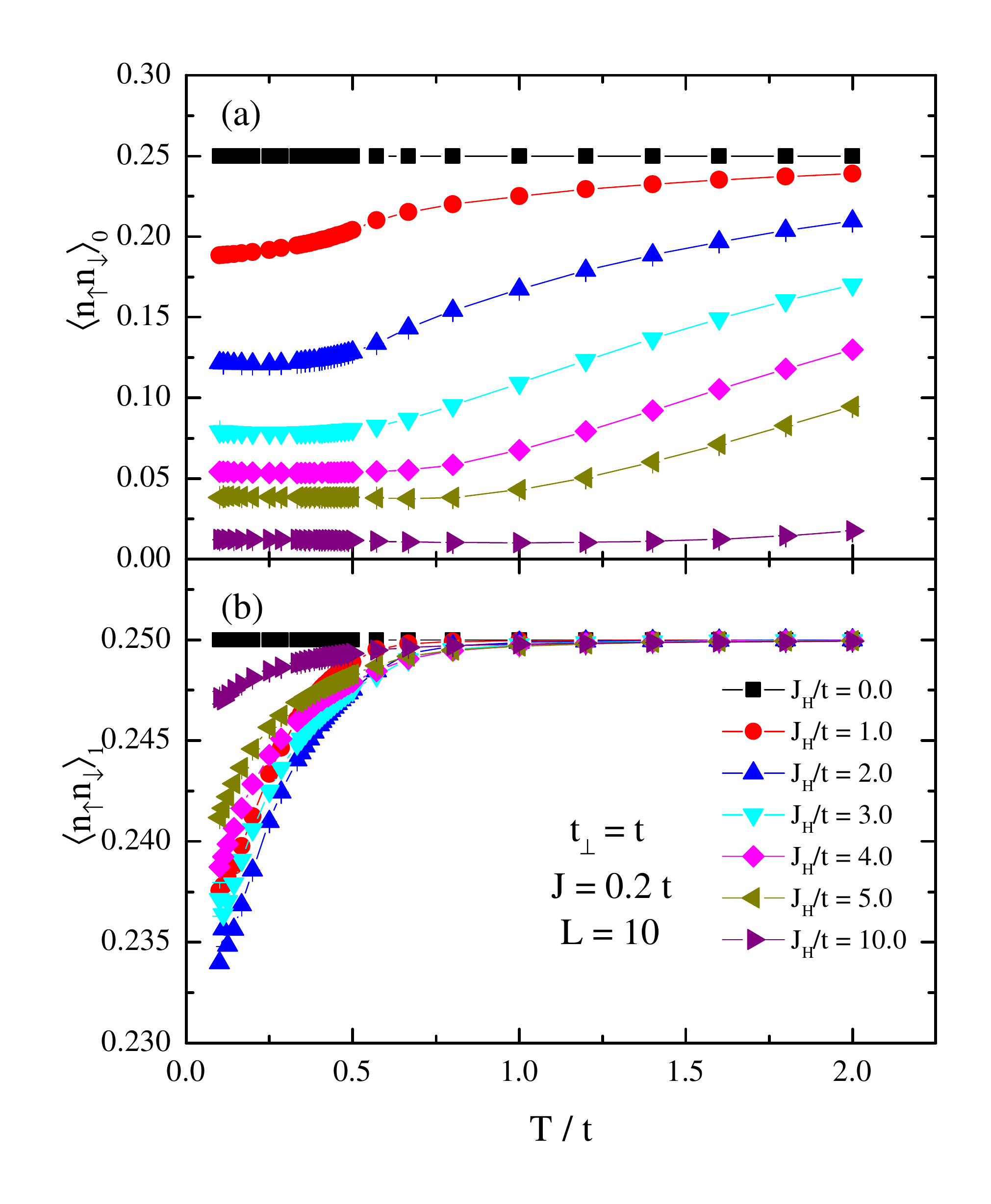,width=9.0cm,angle=-0,clip}
\caption{(Color online)
Temperature dependence of the double occupancy in the case $t_\perp=t$,
in (a) for $\ell=0$, while in (b) for $\ell=1$.
In the former, the increase of $J_H$ decreases double occupancy
since the electrons are strongly coupled to the Ising spins
and, as a consequence, become more localized. 
For $\ell=1$ the effect is very small (note the vertical scale)
a slight decrease in double occupancy and then a recovery 
towards the $J_H=0$ value which begins at $J_H/t \gtrsim 2$.
\label{fig:nudvsT}
}
\vspace{-0.5cm}
\end{figure}  


Having described the effect of the interaction $J_H$ between the Ising
spin plane and the metal on the ordering transition of  the classical
spins, we turn now to the issue of the effect of $J_H$ on the metal.  We
calculate several quantities that characterize both the magnetism and
the transport in the fermionic planes.  We begin by showing, in
Fig.~\ref{fig:kxyvsT}, the intra-plane kinetic energy\cite{footnote3} as
a function of temperature for the different values of $J_H$.  In (a),
which gives the kinetic energy of electrons in the fermionic plane right
at the interface, the increase of the interaction with the Ising spins
localizes the electrons, eventually driving their kinetic energy to 
small values.  This trend is monotonic in $J_H$.  In (b), the farthest
plane from the interface with the Ising spins, we find a much weaker
effect, as is expected in the absence of direct contact with the
classical spin layer.  There is a steady increase of the
absolute value of the kinetic energy- the opposite of the effect
seen in layer $\ell=0$.  The sharp crossover temperature in
the fermion kinetic energy aligns with $T_c$ for the classical spins,
as given in Fig.~\ref{fig:T_cvsJ_H}.

The double occupancy $\langle n_\uparrow n_\downarrow \rangle$ provides
complementary information to the kinetic energy, and in particular
provides insight into the formation of local moments $\langle m^2
\rangle$ and the possibility of Mott metal-insulator behavior.
Specifically, $\langle m^2 \rangle = 1 - 2 \langle n_\uparrow
n_\downarrow \rangle$ at half-filling so that vanishing double occupancy
implies a well-formed moment on every site, and a non-zero double
occupancy implies moments which are partially suppressed by charge
fluctuations.  $\langle n_\uparrow n_\downarrow \rangle(T)$ is shown in
Fig.~\ref{fig:nudvsT}.  Data for plane 0 and plane 1 are shown in the
top and bottom panels, respectively.  In both cases $\langle n_\uparrow
n_\downarrow \rangle$ takes on its uncorrelated value $\langle
n_\uparrow n_\downarrow \rangle =\langle n_\uparrow \rangle \langle
n_\downarrow \rangle = 1/4$ for $J_H=0$, as should be the case for a
metal with no interactions.  In plane 0, there is a monotonic
suppression of double occupancy with $J_H$, and hence a steady
development of local moments.  By the time $J_H = 4$ double occupancy
has decreased to $\langle n_\uparrow n_\downarrow \rangle \sim 0.05$
implying $\langle m^2 \rangle(T) \sim 0.90$.  The reason for this
behavior is clear:  the classical Ising spin $S_i$ acts as a local
magnetic field for the fermions on site $i$ in plane 0,
enhancing(suppressing) the occupation of the electron spin occupation
parallel(antiparallel) to it.  As we shall see, this induced moment
formation aids in magnetic ordering.

\begin{figure}[t]
\epsfig{figure=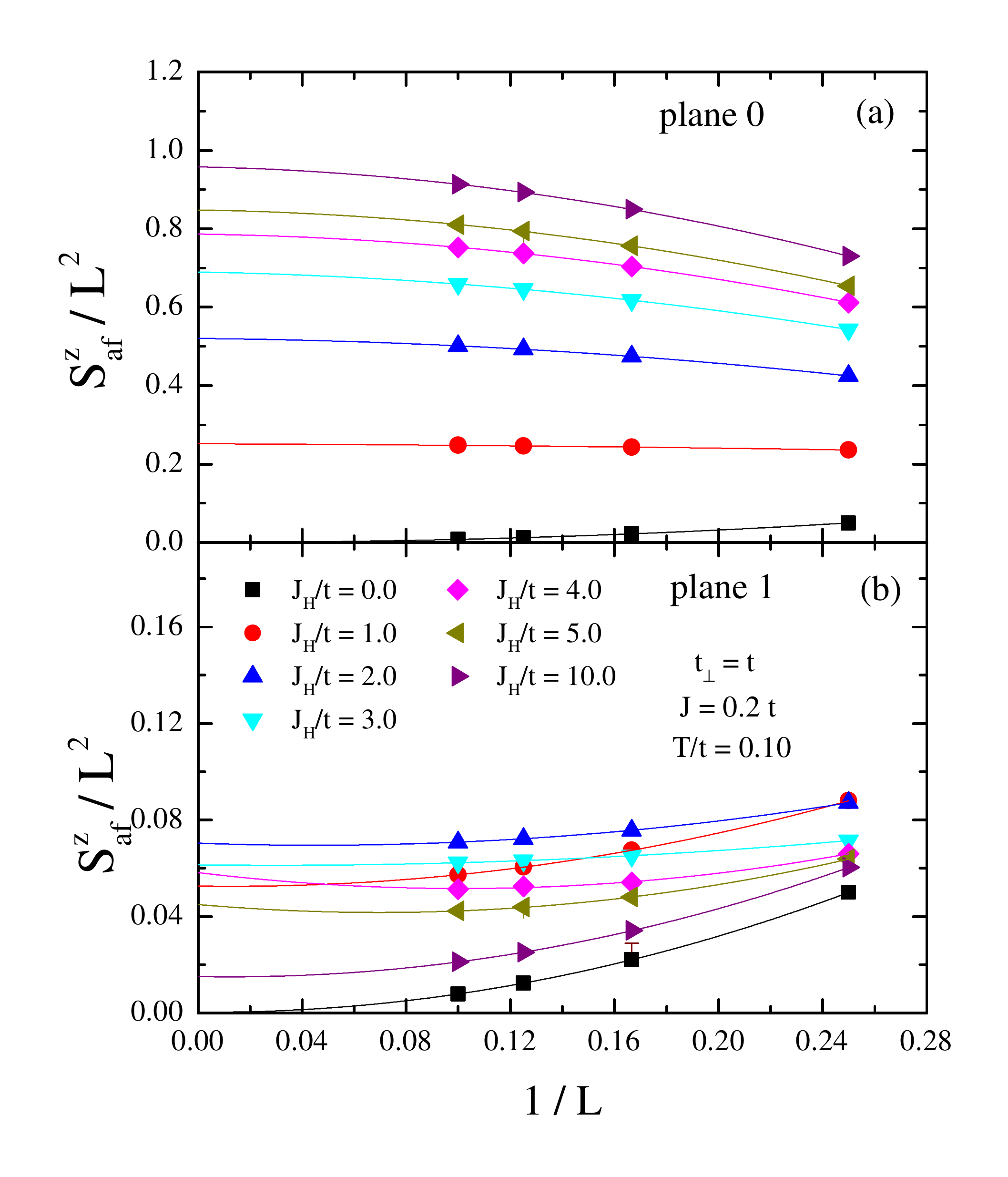,width=9.0cm,angle=-0,clip}
\caption{(Color online)
Finite size extrapolation for the $z$ component of the 
antiferromagnetic structure factor 
$S_{\rm af}^z$ for planes 0 and 1, in (a) and (b), respectively. 
Here $t_\perp=t$ and $T/t = 0.10$.  
A parabolic fit is used to obtain the value in the 
thermodynamic limit, $m_{{\rm af},z}^2$.
\label{fig:safvs1_L}
}
\vspace{-0.5cm}
\end{figure}  

In plane 1, more isolated from the classical spins, the double occupancy
is barely modified from its $J_H=0$ value.  Nevertheless, despite
exhibiting only a small efect, the onset of deviations provides a nice
signal of  the Ne\'el transition temperature.  Indeed, the
non-monotonicity of $T_{\rm N\acute{e}el}$ observed in Fig.~\ref{fig:T_cvsJ_H}
is reflected in a similar non-monotonicity in the double occupancy in
fermionic plane 1.  Presumably, the large response of the double
occupancy to the effective field in plane 0, which is evident far above
$T_{\rm N\acute{e}el}$, masks the more subtle signature of the onset of long
range order.

\begin{figure}[t]
\vspace{-0.5cm}
\epsfig{figure=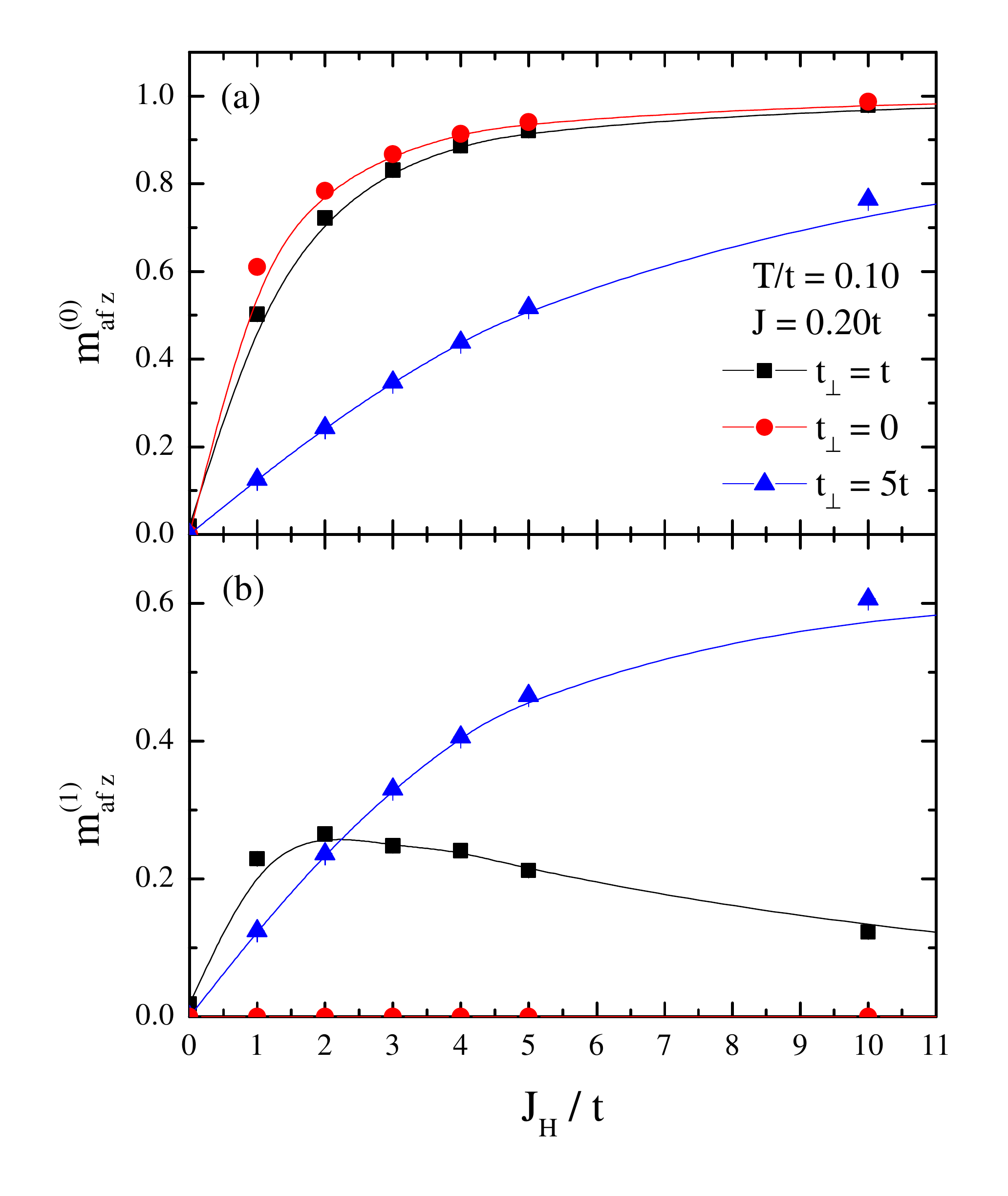,width=9.0cm,angle=-0,clip}
\caption{(Color online)
Antiferromagnetic order parameter as a function of $J_H$
for temperature $T/t =0.10$. In (a) for plane 0 and
in (b) the same for plane 1.
Plane 0 exhibits a rapid and monotonic saturation with $J_H$.
$m^{(1)}_{\rm af\, z}$ in plane 1 first increases with $J_H$ and then falls.
\label{fig:m_AFvsJ_H}
}
\vspace{-0.5cm}
\end{figure}

Long range order of the spin in the metallic planes can be analyzed 
by a finite size scaling of the
antiferromagnetic structure factor,\cite{footnote4}
\begin{equation}
S_{\rm af}^z=\frac{1}{L^2} \,
\sum_{i,j} (-1)^{i+j} \langle s_i^z s_j^z\rangle
= m_{{\rm af},z}^2 + \frac{A}{L} + \frac{B}{L^2}
\label{eq:fss}
\end{equation}
Here $m_{{\rm af},z}$
represents the magnetic order parameter in the metallic layer,
and the sum over $i,j$ is restricted to that same layer.
The coupling of the Ising spins with the $z$ component of the fermionic
spins breaks the $SU(2)$ symmetry of the Hubbard hamiltonian, leading to
the possibility of 
long range order at finite temperature.
Figure~\ref{fig:safvs1_L} shows the extrapolation according to
Eq.~\ref{eq:fss}.  We chose $t_\perp=t$, and separate the contributions
of plane 0 and 1 in (a) and (b), respectively.  However, we do not
attempt to discern this possibility, and restrict ourselves to examining
the ground state magnetism by setting $T=t/10$ where the structure
factor has saturated to its ground state value.  The values of $m_{{\rm af},z}^2$ in the two layers, obtained
from the thermodynamic limit $1/L\rightarrow0$ extrapolation, are
displayed in Fig.~\ref{fig:m_AFvsJ_H} for the same three cases for the
interplane hopping appearing in Fig.~\ref{fig:T_cvsJ_H}. Again, the plot
is separated in (a) and (b) corresponding to the planes 0 and 1,
respectively.  

\begin{figure}[t]
\epsfig{figure=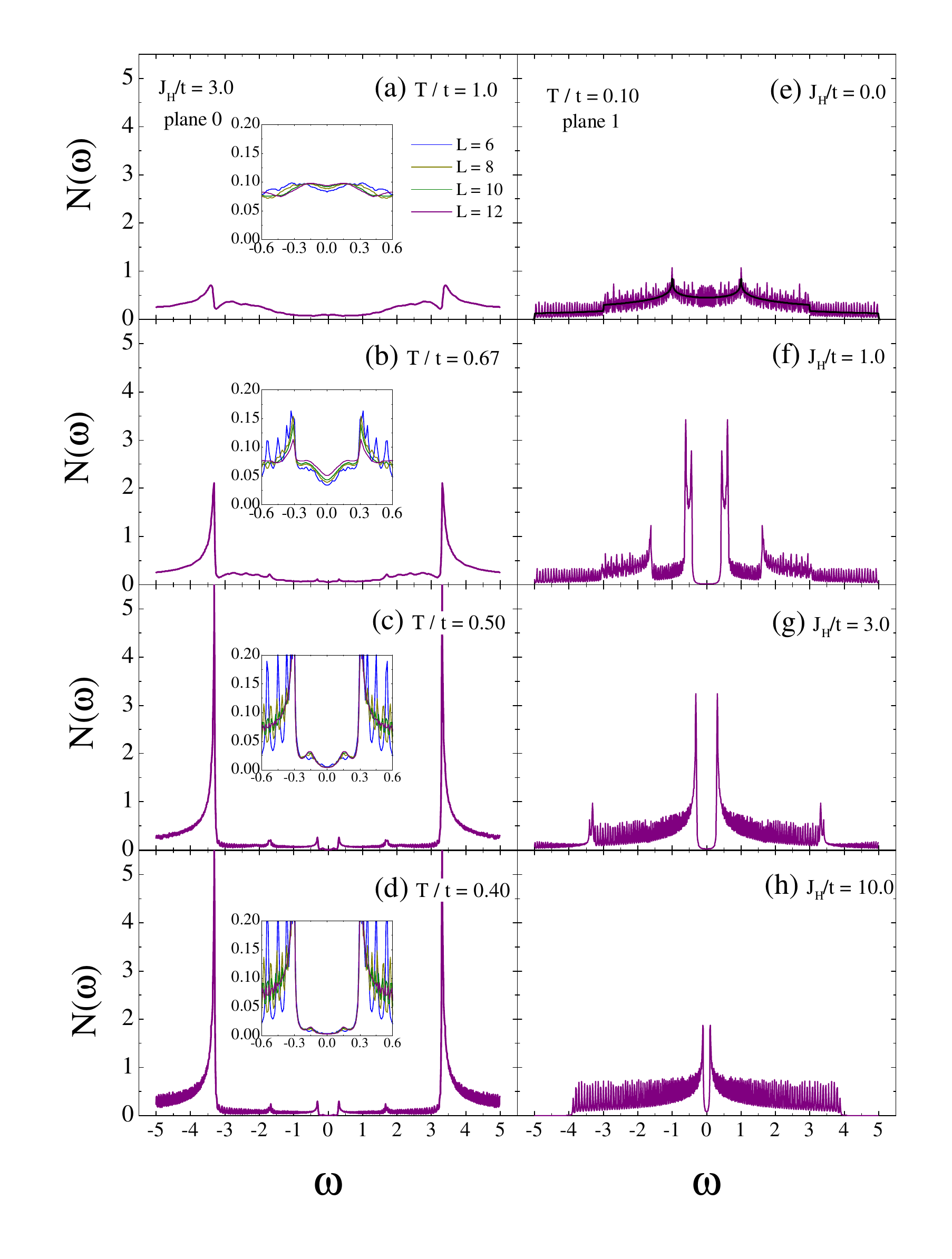,width=9.0cm,angle=-0,clip}
\vspace{-0.5cm}
\caption{(Color online)
\underbar{Left columns:} Density of states $N(\omega)$ of fermions
in plane 0 for different temperatures $T/t=1.00, 0.67, 0.50, 0.40$
(a-d).
The linear lattice size $L=12$, Ising exchange coupling $J=0.2t$,
classical spin-fermion spin coupling $J_H=3t$, and the interplane hopping
$t_\perp=t$.  There are two interesting features in $N(\omega)$:  A
pile-up of density at $\omega \sim \pm J_H$, which is present even at
high $T$, and a gap which opens in the vicinity of $\omega=0$ when $T$ is 
decreased. Insets display the finite size dependence around $\omega=0$.
See text for further
discussion.
\underbar{Right columns:} Density of states $N(\omega)$ of fermions
in plane 1 for different $J_H/t=0.0, 1.0, 3.0, 10.0$ (e-h).
The temperature $T/t = 0.10, t_\perp=t,$ and $L=12$.
A gap is present for finite $J_H$, but gets filled
for larger $J_H$.  (See text for discussion.)
\label{fig:dosAFv2}
}
\vspace{-0.5cm}
\end{figure}  

While plane 0, which directly interacts with the Ising spins, becomes
easily ``saturated" ($m_{{\rm af},z}\rightarrow1$) with the increase of $J_H$,
the fermions on plane 1, farther from the classical spins, are less
easily aligned.  For smaller values of $J_H$, the long range-order
present in the plane at the interface is propagated farther inwards.
However, for $J_H/t \gtrsim 3.0$, the magnetism in plane 1 gets less
robust.
Indeed, the reduction of magnetic order in plane 1
coincides with saturation of magnetic order in plane 0.

We turn now to the issue of how the coupling  to the classical Ising
spins affects the density of states $N(\omega)$ of the metal.  There are
two separate issues to consider.  First, even at high temperatures, the
fluctuating classical Ising spins act as a random site energy $\pm J_H$
for the fermions.  In the limit $t=t_\perp=0$, we expect $N(\omega)=\frac{1}{2}
\big( \, \delta(\omega+J_H) + \delta(\omega-J_H) \, \big)$.  Nonzero
hopping will broaden this distribution.  Second, as $T$ is lowered, the
Ising spins no longer fluctuate randomly but instead, for $J>0$, form an
ordered antiferromagnetic pattern.  This staggered site energy opens a
gap in the fermionic spectrum.
Figure~\ref{fig:dosAFv2} shows $N(\omega)$ for $L=12$ and $t_\perp=t$.
The left panels give $N(\omega)$ in plane 0 for fixed $J_H/t=3$
(the density of states for individual planes is obtained by the
appropriate restriction of the spatial sum in Eq.~\ref{eq:dos}.)
and decreasing temperature.  Both features discussed above are present:  
peaks in the density of states at $\pm J_H$ at all temperatures, and,
near $\omega=0$, an insulating gap which opens only below the Ising
$T_c \sim 0.615$ (Fig.~\ref{fig:Binder1}) for $J_H/t=3$.
The insulating gap is substantially less than one might expect from
a strictly rigid staggered site energy.  Presumably, this
reflects some residual fluctuations of the Ising spins.

In the right panels of Fig.~\ref{fig:dosAFv2} the density of states in the
plane further from the interface is shown.
In the topmost panel Fig.~\ref{fig:dosAFv2}(e), where $J_H=0$,
we recover the analytic result of 
the DOS of a bilayer with $t_\perp=t$ (displayed as a black thick
curve), with some additional structure associated with the discrete
finite lattice peaks.
For $J_H$ nonzero, the antiferromagnetic gap induced in layer 0
propagates to layer 1, rendering it insulating as well.
The size of the gap in $N(\omega)$ for
layer $\ell=1$ goes down for large $J_H$, consistent with the decrease in
the AF order parameter (Fig.~\ref{fig:m_AFvsJ_H}(b)).
One picture of the induced antiferromagnetism, and associated gap,
in the layer not adjacent to the Ising spins, is the following:
when the Ising spins order they induce antiferromagnetism in plane 0
via $J_H$.  It is preferable to have a fermion in plane 1 of
opposite spin from the one above it in plane 0, because it can then 
hop in the perpendicular direction, a lowering of the kinetic energy
which is forbidden by the Pauli principle if the plane 1 fermion
has parallel spin to the plane 0 fermion.

\subsection{The ferromagnetic case}

The antiferromagnetic tendency of tight binding electrons on
a square lattice at half-filling
can be understood from a weak coupling perspective:
The Fermi surface is nested at the antiferromagnetic
ordering vector $(\pi,\pi)$ and, as a consequence, the
non-interacting susceptibility 
\begin{equation}
\chi_0(\mathbf{q})=\frac{1}{L^2}\sum_{\mathbf{k}}
\frac
{f(\epsilon_\mathbf{k})-f(\epsilon_{\mathbf{k} +\mathbf{q}})}
{\epsilon_{\mathbf{k}+\mathbf{q}} -\epsilon_\mathbf{k}}, 
\label{eq:rpa}
\end{equation} 
diverges there as $T \rightarrow 0$.
This reasoning suggests $T_c$ might be suppressed for
ferromagnetically coupled Ising spins, whose ordering wave-vector
conflicts with what the half-filled metallic fermion spins prefer.

\begin{figure}[t]
\epsfig{figure=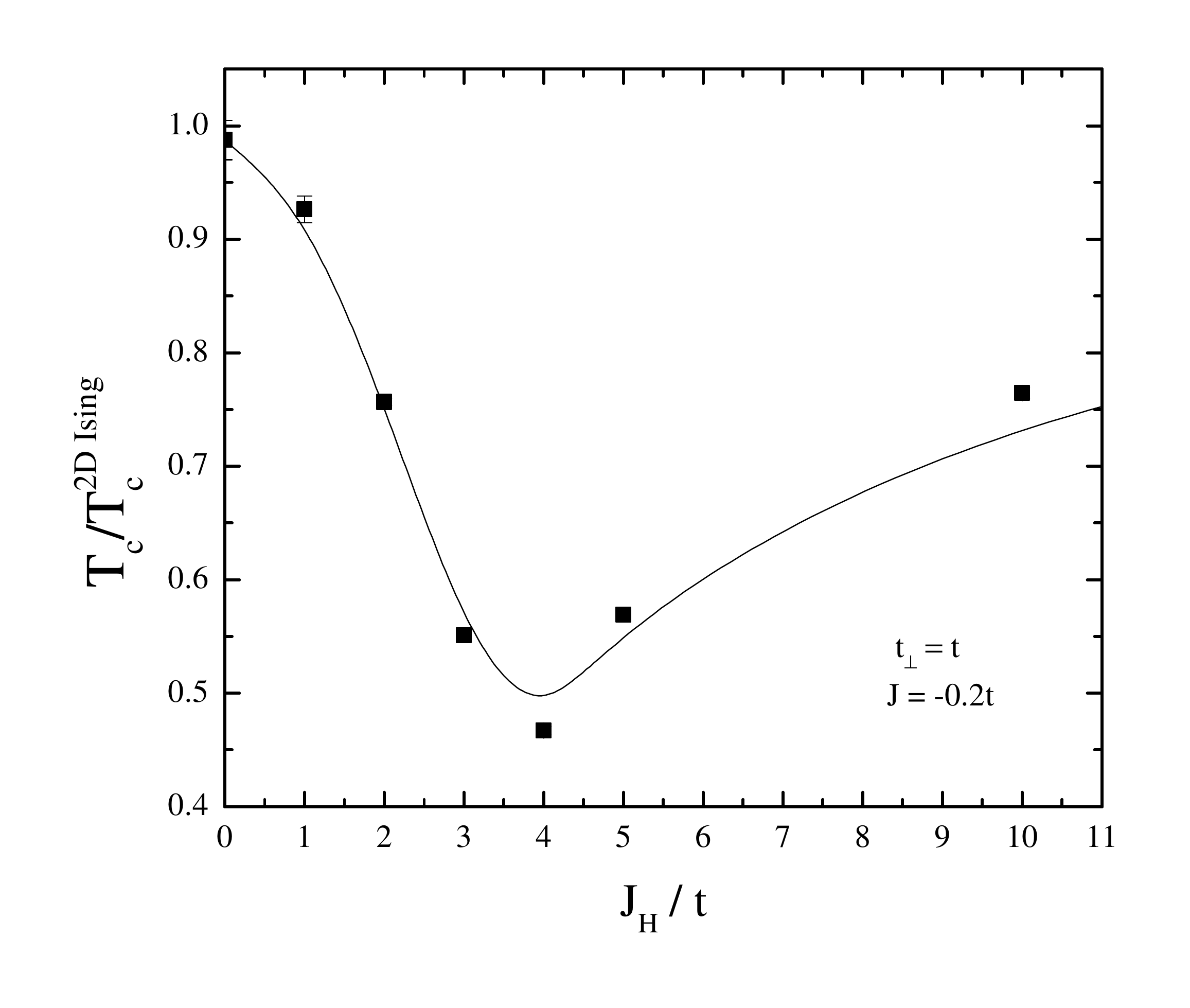,width=9.0cm,angle=-0,clip}
\caption{(Color online)
Curie temperature for the ferromagnetic Ising model with $J=-0.2t$ coupled
with two fermionic planes ($t_\perp=t$),
as a function of $J_H$.
Unlike the antiferromagnetic case, the
coupling with metal decreases the ferromagnetic critical
temperature. Lines are guides to the eye.
\label{fig:Tc_FvsJ_H}
}
\end{figure}

Figure \ref{fig:Tc_FvsJ_H} shows the transition temperature $T_c$ of
ferromagnetically coupled Ising spins in contact with a half-filled
metallic layer.  It confirms that $T_c$ is suppressed, consistent with
the qualitative argument suggested above, and in contrast to the
enhancement seen in the antiferromagnetic case of
Fig.~\ref{fig:T_cvsJ_H}.   The maximal suppression of $T_c$
occurs at $J_H/t \approx 4$, and reveals a lowering
of $T_c$ by almost a factor of two.
$T_c$ ultimately  recovers, but only for very large
values $J_H/t \gtrsim 5$.

\begin{figure}[t]
\epsfig{figure=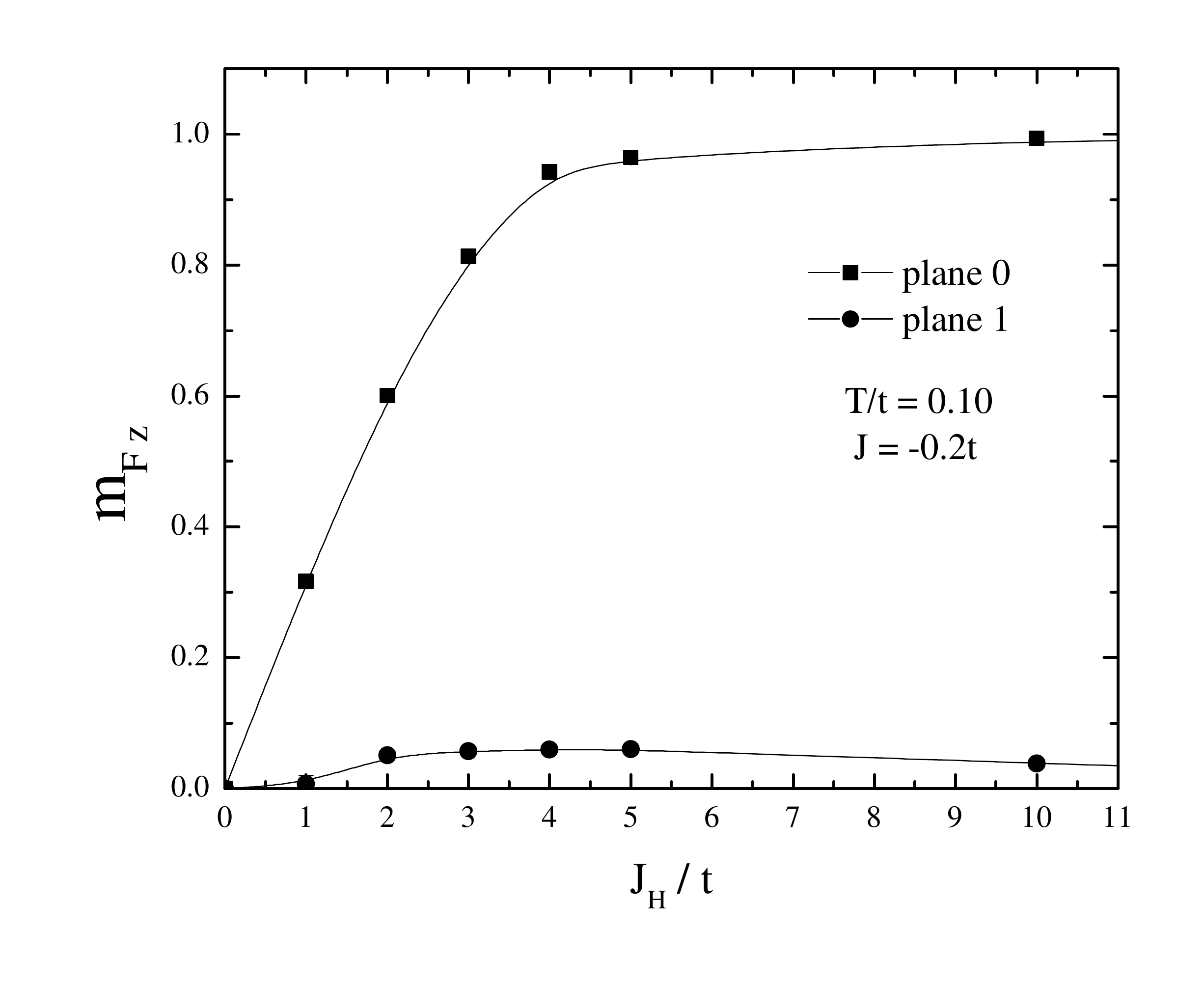,width=9.0cm,angle=-0,clip}
\caption{(Color online)
Dependence of the ferromagnetic order parameter $m_F$
for the itinerant spins as a 
function of the interaction $J_H$ 
with a ferromagnetic Ising plane.  We choose temperature 
$T = t/10$. 
Here $t_\perp=t$ and $J=-0.2t$. 
\label{fig:m_FvsJ_H}
}
\end{figure}  

\begin{figure}[t]
\epsfig{figure=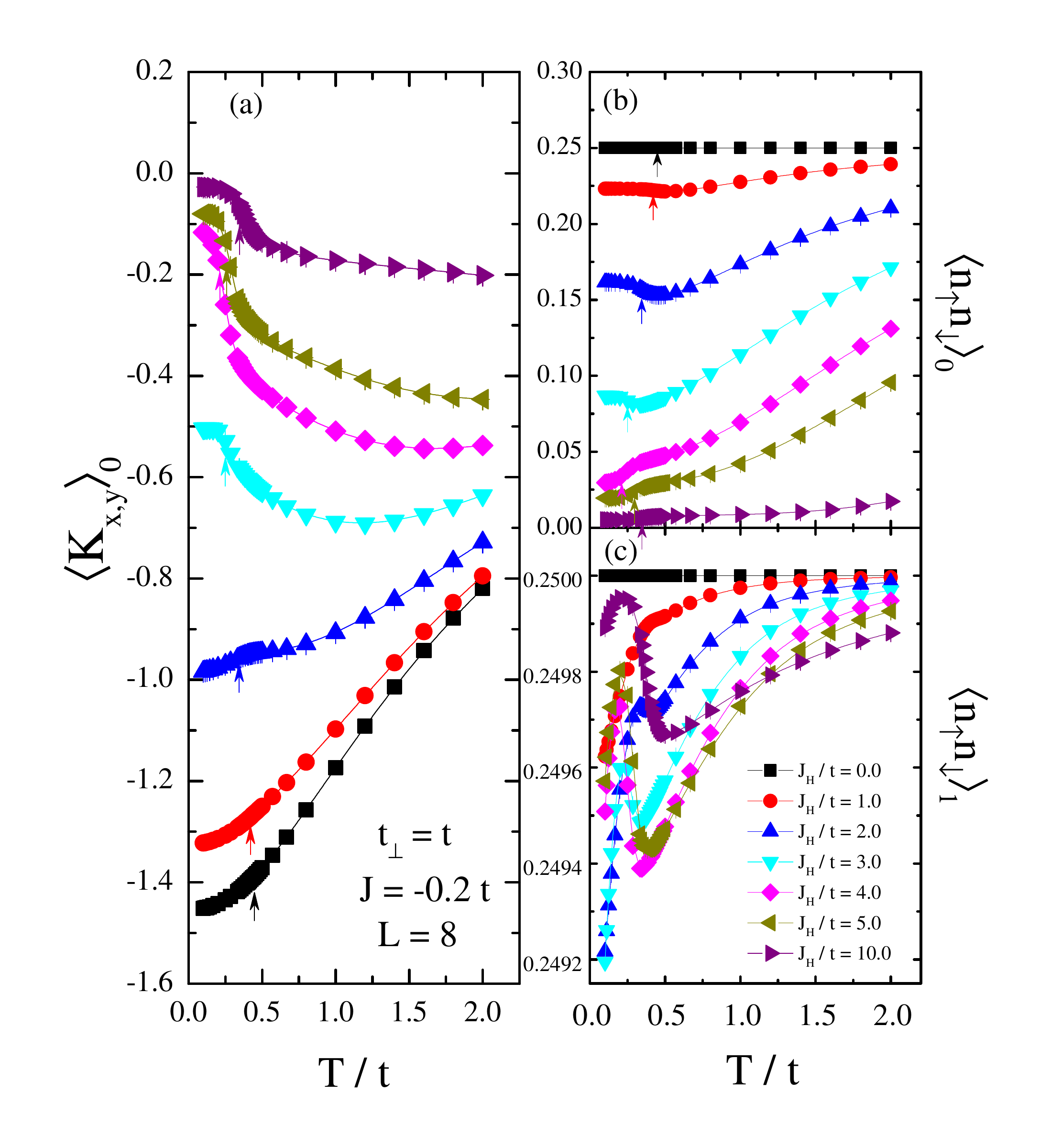,width=9.0cm,angle=-0,clip}
\caption{(Color online)
In (a), temperature dependence of the intra-plane kinetic energy for $\ell=0$ 
in the case $t_\perp=t$ and ferromagnetic interaction between
the Ising spins with exchange constant $J=-0.2t$. $J_H/t\gtrsim3.0$ marks a 
distinct behavior where the kinetic energy decreases rapidly with decreasing 
temperature
in opposition to the cases where $J_H$ is small.
Equivalently to Fig.\ref{fig:nudvsT}, the temperature dependence of the double 
occupancy in planes $\ell=0$ and $\ell=1$
,(b) and (c), respectively. In the former, similar to the antiferromagnetic 
case, the increase in moment localization due to the 
interaction with the neighbor Ising spins can be readily seen. In the 
latter, despite the upturn of double occupancy for low temperatures and large 
interactions, 
the later downturn for even smaller temperatures indicates that the 
ferromagnetism of the Ising layers starts to propagate through the more distant 
fermionic region. Again, in panels (a) and (b), the vertical arrows depict the Ising critical temperature $T_c$.
\label{fig:kxynudvsT_Ferro}
}
\vspace{-0.5cm}
\end{figure}  

When ferromagnetic order is present in layer 0, we can ask whether
it will induce similar order in the more distant layer 1, 
something which occurred with antiferromagnetic coupling $J$.
We calculated the
ferromagnetic structure factor ($S^z_{\rm f}=(1/L^2)\sum_i\langle
s^z_is^z_j\rangle$) for the same values of interaction $J_H$ considered
in the previous case.  A similar finite size analysis,
Eq.~\ref{eq:fss}, was performed, the order parameter $m_F$ for
\emph{ferromagnetism} in each of the fermionic planes was obtained as
function of $J_H$ for low temperature, and is shown in Fig.~\ref{fig:m_FvsJ_H}. 
Ferromagnetic order is induced in both planes, although $m_{\rm f}$ 
is an order of magnitude smaller for plane 1 than for plane 0, in
contrast to the antiferromagnetic case where there was only
a factor of two difference.

As they order,
the fermions in direct contact with the Ising spins ($\ell=0$) start to 
localize as seen in their reduced double occupancy
(Fig.~\ref{fig:kxynudvsT_Ferro}(b)).
For any $J_H$ and $T$,
the double occupancy for $\ell=1$ is basically unchanged from its
uncorrelated value $1/4$ 
(Fig.~\ref{fig:kxynudvsT_Ferro}(c)).
This is similar to what happend in the AF case of 
Fig.~\ref{fig:nudvsT}(b).
The behavior of the $\ell=0$ kinetic energy 
(Fig.~\ref{fig:kxynudvsT_Ferro}(a)),
on the other hand, is quite different from
the AF case (Fig.~\ref{fig:kxyvsT}).
Althougn in both cases there is a systematic suppression with
$J_H$, in the ferromagnetic case the magnitude of the kinetic
decreases as $T$ is lowered for $J_H/t \gtrsim 3$.
This is likely a consequence of the Pauli principle:  In the F
case, ordering of the Ising spins promotes polarization of the
fermions in layer $\ell=0$ and as this polarization becomes
more and more complete the fermions can no longer hop
on the lattice.

\begin{figure}[t]
\epsfig{figure=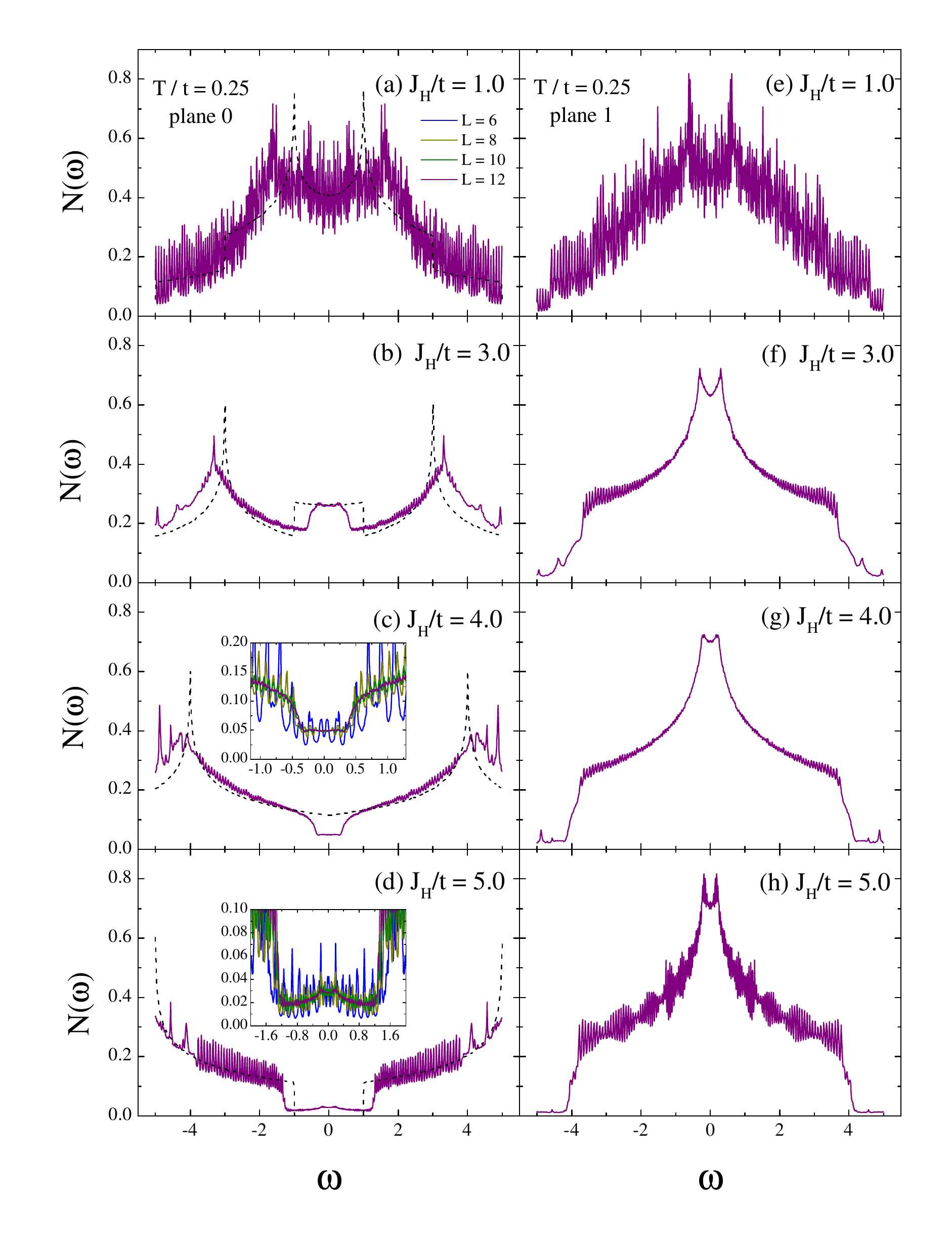,width=9.0cm,angle=-0,clip}
\vspace{-0.5cm}
\caption{(Color online)
Density of states $N(\omega)$ of fermions
in plane $\ell=0$ for different values of the interaction $J_H/t=1.0, 3.0, 4.0, 
5.0$ 
(a-d) at temperature $T/t=0.25$. Panels (e-h) show the correspondent results 
for plane $\ell=1$.
The linear lattice size is $L=12$, Ising exchange coupling $J=-0.2t$ and the 
interplane hopping
$t_\perp=t$. Insets at (c) and (d) include also $L=6,8$ and $10$ near the 
region $\omega=0$. 
Also displayed, as a dashed line, the corresponding density of 
states resulting from the dispersion of one plane under the influence of a 
fixed global chemical potential as if the configuration for the Ising spins is 
``frozen'' in the ferromagnetic state.
Worth noting is that there is  a
pile-up of density at $\omega \sim \pm J_H$, and a pseudogap which opens only 
for 
values of $J_H/t\gtrsim4$. Insets in (c) and (d) show a finite size comparison of 
this gap.
\label{fig:dosFv2}
}
\vspace{-0.5cm}
\end{figure}

Finally, we analyze the influence of the magnetically ordered plane of Ising 
spins on the metallic density of states, Fig.~\ref{fig:dosFv2}.
Similar
to the AF case (Fig.~\ref{fig:dosAFv2}), there are peaks at $\omega \sim 
\pm J_H$ for layer $\ell=0$. The 
increase of $J_H$ induces a pseudogap, however the insets
to (c) and (d) indicate $N(\omega=0)$ remains finite, in contrast to
the AF case.
The dashed line gives the density of states for a 
single fermionic plane coupled to a
perfectly ordered ferromagnetic arrangement for the Ising spins, which is derived from the dispersion $E(\mathbf{k})=-2t(\cos(k_x)+\cos(k_y))\pm J_H$. 
The 
DOS for plane $\ell=1$ is approximately given
by that of a fermionic 
bilayer with $t_\perp=t$.  The effect of $J_H$ 
is to slowly decrease the distance between the van Hove singularities at 
$\omega = \pm t_\perp$. This trend would then ultimately result in a single van 
Hove singularity at $\omega=0$ similar to that of an isolated free 
fermion plane. 
Increasing $J_H$ 
helps ``disconnecting'' the planes $\ell > 0$ which are not right at the
interface.
Similar decoupling can be seen in layered Hubbard models\cite{euverte12}.

\section{Results-  Doped Lattice}
\label{sec:resultsdoped}

In the previous section we analyzed the influence in the critical temperature 
of the Ising plane after attaching a metal to it. Its enhancement(suppression) 
could be explained by the preferred wave-vector of the ordering in this 
metallic region. Since there is a natural tendency for short-ranged 
antiferromagnetic order for free fermions in a tightly-binded bipartite lattice 
at half-filling, these fermions and the Ising spins act 
cooperatively in order to boost the critical temperature of the 
antiferromagnetic Ising model. The same argument shows that when the Ising 
spins have a ferromagnetic coupling, the critical temperature is reduced, once 
again due to the antiferromagnetic tendency introduced by the contact with the 
fermionic spins. 
We now examine the doped lattice where the dominant AF response
in the noninteracting $\chi_0({\bf q})$
of Eq.~\ref{eq:rpa} is no longer present.

\begin{figure}[t]
\epsfig{figure=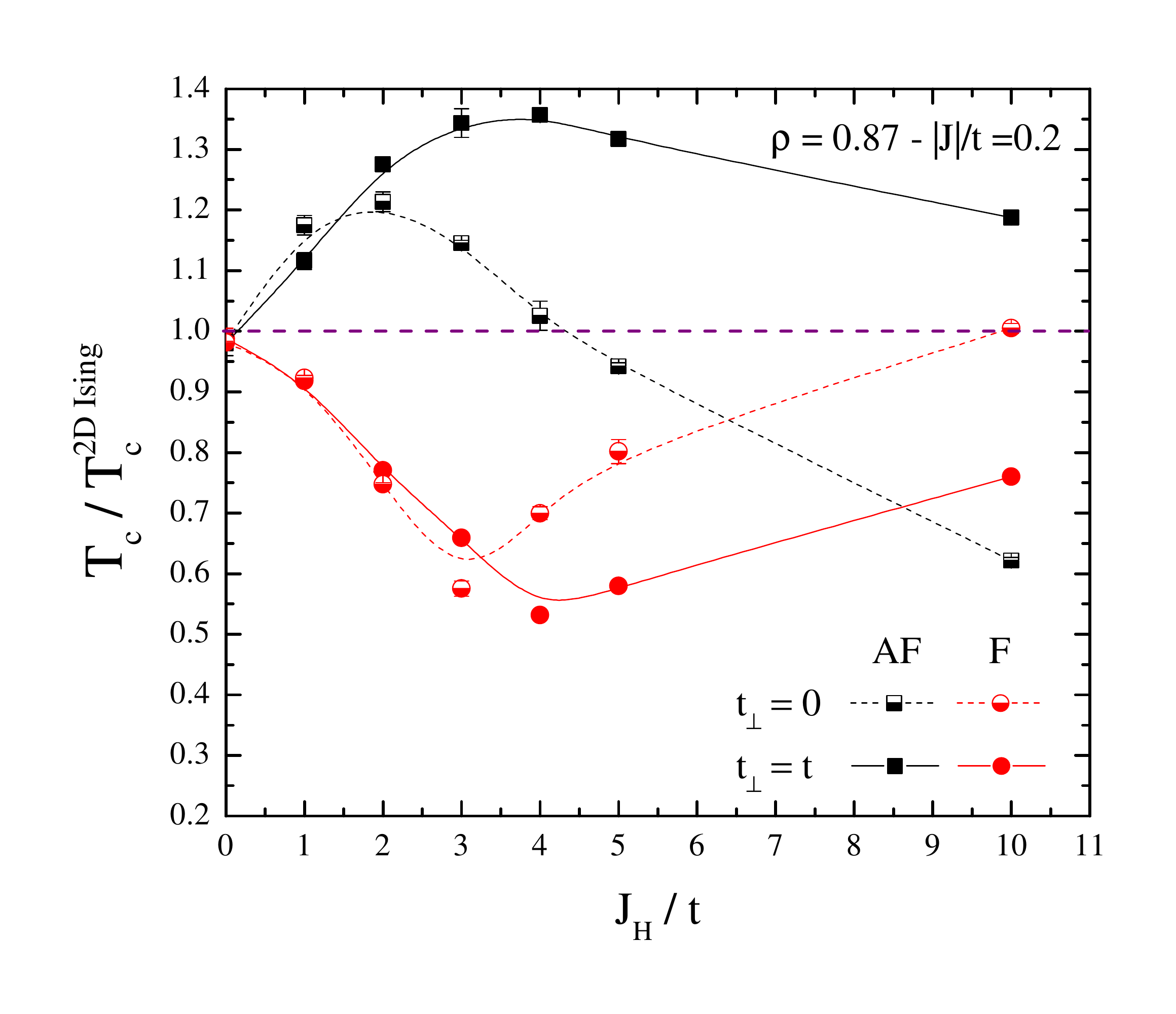,width=9.0cm,angle=-0,clip}
\vspace{-0.5cm}
\caption{(Color online)
Dependence of critical temperature on $J_H$ of the long-range order for the 
Ising spins when coupled to fermions at total density $\rho=0.87$ for 
different scenarios: antiferromagnetic(ferromagnetic) interaction between the 
Ising spins and two fermionic planes coupled by a hopping $t_\perp=t$ and the 
same for the interaction with a single plane.
In the situation one have two fermionic planes, this dependence is 
quantitatively similar to the half-filled case. In the latter scenario, in the 
regime of larger interactions, the coupling with the fermions is detrimental 
(benign) to 
the critical temperature when the antiferromagnetic(ferromagnetic) Ising model 
is considered in a clear contrast with the half-filled case.
\label{fig:T_c_dopedvsJ_H}
}
\vspace{-0.5cm}
\end{figure}

Performing a similar analysis of Fig.~\ref{fig:Binder1} we computed the 
crossings in the Binder ratios for several values of the interaction $J_H$ when 
the metal has a fixed total density $\rho=0.87$. The global chemical potential 
$\mu$ in Eq.~\ref{eq:hamiltonian} is tuned in order to select this density for 
each of the lattice sizes and temperatures values calculated. 
Figure~\ref{fig:T_c_dopedvsJ_H} shows the dependence of the 
critical value of the Ising spins for 
$t_\perp=t$ and $t_\perp = 0$.
When $t_\perp=t$ so that two metallic layers are coupled
to the Ising magnetic layer, the 
qualitative behavior is similar to that at half-filling
(Figs.~\ref{fig:T_cvsJ_H} and 
\ref{fig:Tc_FvsJ_H}).  Indeed, 
the values of the transition temperatures are quantitatively similar.
This is true in both the ferromagnetic and antiferromagnetic
cases. 

However, when $t_\perp=0$, so that 
only one metallic plane is coupled, doping appears to
change the behavior of $T_c$ quite substantially.  While for 
small values of $J_H$ 
the increase(decrease) of the critical temperature of the 
antiferromagnetic(ferromagnetic) Ising model is the same as for
$\rho=1$, once higher values 
of $J_H$ are reached ($J_H/t \sim 4$ in the AF case and $J_H/t \sim 10$ 
in the F one) the scenario 
changes.  An antiferromagnetic Ising plane has its critical temperature 
decreased by the coupling with the free electron spins while in the 
ferromagnetic case the critical temperature is enhanced. 
This is not completely unexpected since, as commented earlier,
the peak in $\chi_0({\bf q})$ moves away from $(\pi,\pi)$ so
that the fermions in the metal  no longer so strongly favor AF order.


\begin{figure}[t]
\epsfig{figure=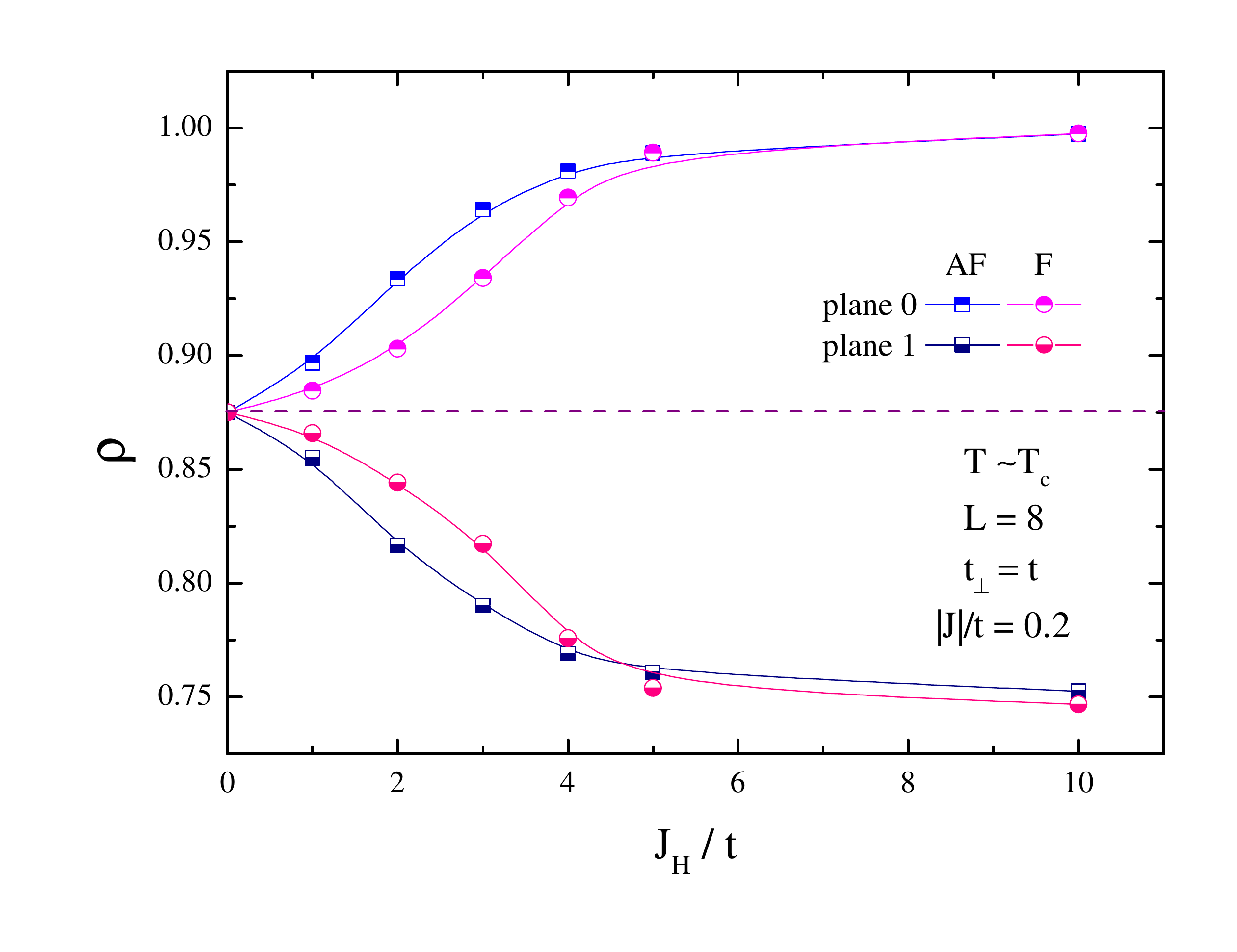,width=9.0cm,angle=-0,clip}
\vspace{-0.5cm}
\caption{(Color online)
Dependence of the density on $J_H/t$ in each of the fermionic planes at temperatures close to where the magnetic transition takes place for the 
Ising spins. The lattice size is $L=8$, $t_\perp = t$ and the interaction among 
the Ising spins are either ferro or antiferromagnetic with value $|J|=0.2t$.
\label{fig:rho_vs_J_H_doped}
}
\vspace{-0.5cm}
\end{figure}

The reason that this does not happen in the two layer case $t_\perp=1$
is that the second metallic layer $\ell=1$ acts as a charge reservoir
for the layer $\ell=0$ at the interface.  That is, the 
electron density is imbalanced, 
as seen in Fig.~\ref{fig:rho_vs_J_H_doped}. 
Plane $\ell=0$ adjacent 
to the magnetic layer has a tendency to become half-filled,
leaving the farthest plane less populated. 
For larger values of $J_H$
the occupations tend to $1.0$ and $0.75$, for $\ell=0$ and $\ell=1$
respectively. 
Throughout this evolution
the total density is preserved at $\rho=0.87$. The 
half-filling of layer $\ell=0$
allows for the
enhancement(supressing) of the critical temperature of an 
antiferromagnetic(ferromagnetic) aligned Ising plane.

\section{Conclusion}
\label{sec:conclusion}

We studied magnetic order at the interface between an insulator and a
metal using quantum Monte Carlo.  Specifically, we considered a 2D Ising
plane coupled to a lattice of noninteracting (metallic) fermions.  In
the case of an antiferromagnetic Ising model, and a half-filled metal,
the coupling enhanced the Ising critical temperature.
Antiferromagnetic order was also induced in the metal, both in the layer
immediately at the interface with the classical spins and also deeper
within.  This enhancement occurs even in the case where the interlayer
hopping in the fermionic sheet is made large enough that the fermions
become a band insulator, namely a bilayer with interplane hopping
$t_\perp$ bigger that $4t$. 

In contrast, the critical temperature of ferromagnetic Ising spins
is reduced by the coupling to the fermions at half-filling.
We attribute these distinct effects to be a consequence of
the perfect nesting of the square lattice fermion tight binding
hamiltonian, which favors antiferromagnetism.
Indeed, studies of the doped lattice demonstrate that the system's
desire to optimize `magnetic consistency', that is to have an AF
metallic response when the magnetic layer is AF, is so great that,
if available, charge will be pulled from a second magnetic
layer into the interface metallic layer so that half-filling
is maintained there.
In the absence of such a reservoir, the AF transition is suppressed
by this mismatch with the metallic ordering vave vector.

A central consideration of our work has been the 
consistency of the order in the classical spin plane from
the ordering tendency of the metal.  In ``unfrustrated"
cases where the metal and local spins prefer the
same wave-vector, transition temperatures are enhanced, and
{\it vice versa}.
Recent experiments 
\cite{langridge14}
have explored the importance of these
considerations on the decoupling of surface and bulk magnetism
in UO$_2$.  The distinct surface behavior observed is attributed to
the different symmetry of its ordered phase relative to the bulk.
Other 3D systems in which 2D order occurs due to frustration are
certain of the doped cuprate superconductors  \cite{tranquada12,
tranquada13}.  In
individual CuO$_2$ sheets, stripes of $d$-wave order coexist with
intervening antiferromagnetic stripes.  The orientation of the $d$-wave
phases alternates from stripe to stripe in a given layer.  
In adjacent CuO$_2$ sheets,
the same stripe pattern occurs, but, because of structural effects, the
stripes are oriented perpendicular to the neighboring sheet.  
The result is that the 
intersheet Josephson coupling tends to cancel and 2D superconductivity
is observed.

A natural progression of the work reported here
would be to consider the case of continuous
XY (planar) or Heisenberg spins.  As previously noted, in this
case an isolated 2D spin plane has no
transition to long range order, owing to the Mermin-Wagner theorem.  
One interesting question will be how the less robust 
power law correlations which develop at low $T$ in the XY
case are qualitatively affected by coupling to the metal.
For $J<0$, where we find the Ising $T_c$ suppressed,
will the Kosterlitz-Thouless transition survive?
Because the fermion determinant depends only on the spin degrees
of freedom in the interface layer, adding additional spin layers 
has relatively little computational cost.
Thus it is feasible to study a 3D lattice Heisenberg spins, which has 
a finite ordering temperature, coupled to one of more 
metallic layers.

\vskip0.15in
{\bf Acknowledgements:}
This work was supported by DOE DE-NA0001842-0, and by the Office
of the President of the University of California.  
Support from  CNPq (TP and 
RM), FAPERJ (TP),
and the CNPq Science Without
Borders project (RS)
is gratefully acknowledged.



\begin{thebibliography}{100}

\bibitem{troyer96}
M. Troyer, H. Kontani, and K. Ueda,
Phys. Rev. Lett. 76, 3822 (1996).

\bibitem{sandvik94}
A.W. Sandvik and D.J. Scalapino,
Phys. Rev. Lett. 72, 2777 (1994).

\bibitem{scalettar94}
R.T.~Scalettar, J.W.~Cannon, D.J.~Scalapino,
and R.L.~Sugar, Phys.~Rev.~B50, 13419 (1994).

\bibitem{euverte12}
A. Euverte, F. Hebert, S. Chiesa,
R.T.~Scalettar, and G.G.~Batrouni,
Phys. Rev. Lett. 108, 246401 (2012).

\bibitem{dagotto01}
E. Dagotto, T. Hotta, and A. Moreo, Phys. Rep. 344, 1 (2001).

\bibitem{lv10}
W. Lv, F. Kr\"uger, and P. Phillips, Phys. Rev. B82,
045125 (2010).

\bibitem{yin10}
W.-G. Yin, C.-C. Lee, and W. Ku, Phys. Rev. Lett. 105,
107004 (2010).

\bibitem{liang12}
S. Liang, G. Alvarez, C. Sen, A. Moreo, and E. Dagotto,
Phys. Rev. Lett. {\bf 109}, 047001 (2012).

\bibitem{takahashi01}
K. S. Takahashi, M. Kawasaki, and Y. Tokura,
Appl. Phys. Lett. {\bf 79}, 1324 (2001).

\bibitem{freeland10}
J. W. Freeland, J. Chakhalian, A. V. Boris, J.-M. Tonnerre, J. J. Kavich, P. Yordanov, S. Grenier, P. Zschack, E. Karapetrova, P. Popovich, H. N. Lee, and B. Keimer,
\prb {\bf 81}, 094414 (2010).

\bibitem{tan13}
H. Tan, R. Egoavil, A. B\'ech\'e, G. T. Martinez, S.VanAert, J. Verbeeck, G. VanTendeloo, H. Rotella, P. Boullay, A. Pautrat, and W. Prellier,
\prb {\bf 88}, 155123 (2013).

\bibitem{dagotto14}
A. Mukherjee, N. D. Patel, S. Dong, S. Johnston, Adriana Moreo, and Elbio Dagotto,
arXiv:1409.6790.

\bibitem{footnote1}
We focus here in the penetration of the classical spin order into
the metal, and hence have a single spin layer and multiple metallic
layers.

\bibitem{chiesa13}
A detailed discussion of mean field treatments and the role of
rotationally symmetric decouplings 
in the context of layered Hubbard models is contained in
J. Xu, S. Chiesa, E.J. Walter, and S. Zhang,
J. Phys. Cond. Mat. 25, 415602 (2013).

\bibitem{zhang88}
F.C. Zhang and T. M. Rice, Phys. Rev. B37, 3759 (1988); 
C. Gros, R. Joynt, and T.M. Rice, {\it ibid.} 36, 381 (1987).

\bibitem{footnote2}
This real-valued matrix turns into a complex one once we define the complex 
phases in the hopping terms to reduce the finite size effects.

\bibitem{motome99}
Y. Motome and N. Furukawa, J. Phys. Soc. Japan 68, 3853 (1999).

\bibitem{alvarez05}
G. Alvarez, C. Sen,
N. Furukawa, Y. Motome, and E. Dagotto,
Comp. Phys. Comm. 168, 32 (2005).

\bibitem{cen06}
C. Sen, G. Alvarez, Y. Motome, N. Furukawa, I. A. Sergienko, T. C.
Schulthess, A. Moreo, and E. Dagotto, Phys. Rev. B73, 224430 (2006).

\bibitem{Binder81}
K. Binder, Zeitschrift f\"ur Physik B Condensed Matter 43, 119
(1981).

\bibitem{gammel92}
J. Tinka Gammel, D. K. Campbell, and E. Y. Loh,
arXiv:cond-mat/9209026.

\bibitem{gros96}
C. Gros, Phys. Rev. B53, 6865 (1996).

\bibitem{chiesa08}
S. Chiesa, P.B. Chakraborty,
W.E. Pickett, and R.T. Scalettar,
Phys. Rev. Lett. 101, 086401 (2008).


\bibitem{Assaad} F. F. Assaad, in {\it Quantum Simulations of Complex 
Many-Body Systems: From Theory to Algorithms}, edited by J. Grotendorst,
D. Marx, and A. Muramatsu (John von Neumann Institute for Computing 
(NIC), 2002), Vol.\ 10, pp. 99-155.

\bibitem{footnote3}
The kinetic energy does not vanish even in the Mott insulator
owing to quantum fluctuations.
See, for example,
S.R. White, D.J. Scalapino, R.L. Sugar, E.Y. Loh, 
J.E. Gubernatis, and R.T. Scalettar,
Phys. Rev. B40, 506 (1989);  and
C.N. Varney, C.R. Lee, Z.J. Bai, S. Chiesa, M. Jarrell, and R.T. Scalettar,
Phys. Rev. B80, 075116 (2009).  



\bibitem{footnote4}
Finite size scaling of the farthest distance spin correlation
can also be done, and leads to the same value for the
order parameter.  In general the structure factor has the advantage
of smaller statistical error bars, but the disadvantage of
larger finite size effects (a bigger value of the parameter $A$ in
Eq.~\ref{eq:fss}).

\bibitem{langridge14}
S. Langridge, G. M. Watson, D. Gibbs, J. J. Betouras, N. I. Gidopoulos,
F. Pollmann, M. W. Long, C. Vettier, and G. H. Lander,
Phys. Rev. Lett. 112, 167201 (2014).

\bibitem{tranquada12}
J.M. Tranquada,
Physica B407, 1771 (2012).

\bibitem{tranquada13}
Z. Stegen, Su Jung Han, Jie Wu, A K. Pramanik, M. H\"ucker,
Genda Gu, Qiang Li, J.H. Park, G.S. Boebinger, and J.M. Tranquada,
Phys. Rev. B87, 064509 (2013).

\end{thebibliography}
\end{document}